# LES and finite-volume CMC modelling of a turbulent lifted $H_2/N_2$ flame: effects of CMC mesh resolution and numerical scheme

Guangze Li[1,2], Huangwei Zhang[2,*], Longfei Chen[1]

[1]School of Energy and Power Engineering, Beihang University, Beijing 100191, China
[2]Department of Mechanical Engineering, National University of Singapore, 9 Engineering Drive 1, Singapore 117576, Republic of Singapore

## Abstract

Large eddy simulations with three-dimensional finite-volume Conditional Moment Closure (CMC) model are performed for a hydrogen / nitrogen lifted flame with detailed chemical meachanism. The emphasis is laid on the influences of mesh resolution and convection scheme of finite-volume CMC model on predictions of reactive scalar distribution and unsteady flame dynamics. The results show that the lift-off height is underestimated and the reactive scalars (e.g. temperature, $H_2$ and OH) are over-predicted with coarser CMC mesh. It is also found that further refinement of the CMC mesh would not considerably improve the results. The time sequences of the most reactive and stoichiometric OH mass fractions indicate that finer CMC mesh can capture more unsteady details than coarser CMC mesh. Moreover, the coarse CMC mesh has lower conditional scalar dissipation rate, which would promote the ealier auto-ignition of the flame base. Besides, the effects of the convection schemes in the CMC equations on the lifted flame characteristics are also investigated. It is shown that different convection schemes lead to limited differences on the time-averaged temperature, mixture fraction and species mass fractions. Moreover, the RMS values of $H_2$ and OH mass fractions show larger deviation from the measurements with hybrid upwind and central differencing scheme, especially around the flame base. Furthermore, the distributions of the numerical flux on the CMC faces also show obvious distinction between the upwind scheme and the blending scheme. The budget analysis of the individual CMC terms shows that a sequence of CMC faces has comparable contributions with upwind scheme. However, with the hybrid schemes, the instantaneous flux is dominantly from limited CMC faces. The reactivity of a CMC cell is more easily to be affected by its neighbors when the upwind scheme is used.

---

[*] Corresponding author. E-mail: huangwei.zhang@nus.edu.sg. Tel: +65 6516 2557.

# 1. Introduction

Turbulent lifted flames have been extensively adopted in practical combustion apparatus, including industrial burner and gas turbine combustor, since they can prevent the nozzle from being damaged by the high-temperature zone [1-3]. Meanwhile, the fuel and oxidant can achieve a certain level of pre-mixing through diffusion and turbulent motion ahead of the flame base. It is generally accepted that lift-off, blow-off and stabilization of the flame are affected by the mutual interaction of turbulence and chemical kinetics [4]. Clear understanding of the complex mechanism behind the lifted flame is of great importance to design and improve the industrial combustion devices.

Numerous experimental studies on turbulent lifted flames have been performed. For instance, Cabra et al. [5] designed a vitiated co-flow burner to explore the features of lifted turbulent $H_2/N_2$ jet flames, and they measured the temperature and key species concentrations in the lifted flame. Using the same burner as in Ref. [5], Wu et al. [6] investigated the correlation between the lift-off height and various flow conditions (e.g. jet / co-flow velocities and co-flow temperature). Their results reveal that the lift-off height increases with jet / co-flow velocities and decreases with co-flow temperature. Moreover, Markides and Mastorakos [7] experimentally studied the autoigniton behavior of hydrogen in a co-flowing air stream, and they reported the similar correlations between the flame lift-off height and co-flow velocity and temperature to those by Wu et al. [6]. In addition, Leung and Wierzba [8] further studied the co-flow velocity effects on stability of turbulent non-premixed jet flame and they found that the co-flow velocity considerably influences the blowout limits of the lifted flames.

Due to the simple flow configurations and well-defined boundary conditions, turbulent lifted flames are widely used for combustion model validations. The Conditional Moment Closure (CMC) approach has been shown to be able to accurately predict the turbulent lifted flames [9-17]. For instance, the LES−CMC approach was used to simulate a lifted methane flame by Navarro-Martinez et al. [15], and their results show that the flow characteristics and reactive scalars are predicted well by the LES−CMC model and the effects of inflow turbulence on lift-off height were also captured satisfactorily. They used the same model to investigate various lifted hydrogen flames [14], i.e. Berkeley experiments [5] and Cambridge experiments [7], which further corroborate the prediction accuracy of the sub-grid scale CMC model. Moreover, Stankovic [13] simulated hydrogen auto-ignition in a turbulent co-flow of heated air also with LES−CMC approach, and various experimentally observed autoignition regimes are reproduced by LES−CMC. With LES−CMC, Tyliszczak [16] assessed the effects of different models of conditional scalar dissipation rate on auto-ignition of lifted hydrogen flame, and it is shown that the predicted lift-off height is sensitive to the model constant for sub-grid scale scalar dissipation. Rosiak and Tyliszczak [17] studied the flame development and propagation of a pure hydrogen jet in a hot co-flow of oxygen and water vapor with LES−CMC approach, and found that the changes of the oxidizer composition can impact the maximum flame temperature and lift-off height.

In the above LES-CMC simulations [9-17], the CMC equations are discretized with finite differencing method on a different mesh from the LES one. Recently, to achieve higher prediction accuracy of the physical transport terms and accommodate more realistic turbulent flame problems (e.g. model gas turbine combustors), the LES-CMC model based on finite volume discretization is implemented, extended from the previous Cambridge finite-differencing solver [18-20]. The essence of this implementation is to discretize the CMC equations (more specifically physical transport terms and relevant quantities) based on the surface fluxes through the CMC cell faces and these CMC faces are automatically selected from the cell faces of fine LES mesh. The improvements for the CMC simulations include: 1) Polyhedral CMC cells can be used, rendering it suitable for complicated flame configurations (e.g. model burners and/or variable inlet conditions). This is an important step for an advanced combustion model for real applications. 2) Since the CMC faces are selected from the LES faces, the numerical fluxes are essentially resolved at the (fine) LES resolution, instead of the (coarse) CMC resolutions. Therefore, the variations (e.g. fluctuations) of the surface fluxes for a CMC cell can be accurately captured, compared to the numerical discretization done over the coarse CMC resolutions. 3) At the inlet conditions, due to the surface flux calculations based on LES (or CMC) faces, the inlet condition effects (e.g. inlet turbulence) on the near-inlet CMC cells are accurately quantified. It has been validated in predicting localized / global extinctions and forced ignition in turbulent non-premixed flames [18, 21-24]. However, whether the finite volume CMC model can accurately predict the auto-ignition of turbulent lifted flames has not been examined yet. Clarifying this would be helpful to extend the finite volume CMC model for more complicated problems, e.g. turbulent lifted spray flames.

Moreover, despite the successful applications of the finite-volume CMC model [18, 19, 21-24], the sensitivity to the model implementations has not been particularly investigated. It is known that due to distinct LES and CMC meshes, the CMC resolution may affect the calculations of reactive scalars in mixture fraction space and strong spatial variation of flame structure in physical space. Stankovic et al. [12] used LES with the finite differencing CMC to investigate the Cabra lifted $H_2/N_2$ flame [5], and they found that the refinement of CMC mesh may move the flame base further downstream. Navarro-Martinez and Kronenburg also analyzed various CMC resolutions in their LES of different lifted flames with one- and two-dimensional CMC models [14], and observed that the coarser CMC mesh would produce an unrealistically flat temperature profile at the flame base. Therefore, whether the conclusions from the above work [12, 14] can be straightforwardly extended to the three-dimensional finite volume CMC model is uncertain. Besides, the interactions between the neighboring CMC cells (characterized by the convection and sub-grid scale diffusion of the conditional reactive scalars) have significant effects on capturing highly unsteady and localized flame dynamics (e.g. localized extinction, blow-off and re-ignition) [14, 21]. In Refs. [18, 19, 21-24], the upwind scheme is used for the convection term in the CMC equations. With it, the transport direction of conditional reactive scalars in physical space is dominantly controlled by the local mass or volume flux. Whether the dissipative nature of this scheme affects the predictions of instantaneous and/or local flame dynamics needs to be assessed. It is worth mentioning that both the numerical schemes and the ratios of CMC and LES filter sizes are numerical parameters that may influence the prediction accuracies of the LES-CMC model.

In this work, the effects of CMC resolution and numerical scheme will be studied. The lifted $H_2/N_2$ flame [5] is selected as the target flame. The reasons can be explained as: 1), The uncertainties of hydrogen chemical mechanism is relatively small; 2), There are plenty of experimental data measured from the Cabra flame for model validations; 3), The richness of unsteady flame dynamics, such as flame autoignition and stabilization as well as turbulence-chemistry interaction, is helpful for examining the prediction ability of the LES-CMC model. In this work, three CMC meshes are adopted to investigate the resolution sensitivity in the framework of finite volume discretization. It is worth mentioning that the finest CMC mesh studied in this work is identical to the LES mesh, which is helpful to assess the accuracies of the data exchange between the LES and CMC meshes. Furthermore, three convection schemes, including upwind scheme, hybrid upwind and central differencing schemes with different blending factors, are employed to explore the discretization scheme effects.

The rest of the manuscript is structured as below. Section 2 presents the governing equations for LES and CMC modelling. The flame information and numerical implementation are introduced in Section 3, followed by the results and discussion in Section 4. The main conclusions are summarized in Section 5.

## 2. Governing equation

2.1 *Large eddy simulation*

In LES, the large-scale eddies are resolved, while the effects of the unresolved small-scale eddies on the resolved flow field are modelled. Their governing equations can be derived through low-pass filtering the respective instantaneous equations. In this work, the resolved continuity and momentum equations are solved [25]

$$\frac{\partial \bar{\rho}}{\partial t} + \frac{\partial \bar{\rho}\tilde{u}_i}{\partial x_i} = 0, \tag{1}$$

$$\frac{\partial \bar{\rho}\tilde{u}_i}{\partial t} + \frac{\partial \bar{\rho}\tilde{u}_i\tilde{u}_j}{\partial x_j} = -\frac{\partial \bar{p}}{\partial x_i} + \frac{\partial \tilde{\tau}_{ij}}{\partial x_j} - \frac{\partial \tau_{ij}^{sgs}}{\partial x_j}, \tag{2}$$

where $t$ is time, $x$ is spatial coordinate, $\bar{\rho}$ is resolved density, $\bar{p}$ is resolved pressure and $\tilde{u}$ is resolved velocity. $\tilde{\tau}_{ij} = \mu(\frac{\partial \tilde{u}_i}{\partial x_j} + \frac{\partial \tilde{u}_j}{\partial x_i} - \frac{2}{3}\frac{\partial \tilde{u}_k}{\partial x_k}\delta_{ij})$ is the resolved stress tensor with $\delta_{ij}$ being Kronecker delta function. $\mu$ is the dynamic viscosity, which is estimated based on Sutherland's law. $\tau_{ij}^{sgs} = \bar{\rho}(\widetilde{u_i u_j} - \tilde{u}_i \tilde{u}_j)$ is the sub-grid scale stress tensor, closed by the constant Smagorinsky model [26].

For modelling turbulent non-premixed flames, the resolved mixture fraction $\tilde{\xi}$ is solved from

$$\frac{\partial \bar{\rho}\tilde{\xi}}{\partial t} + \frac{\partial \bar{\rho}\tilde{\xi}\tilde{u}_j}{\partial x_j} = \frac{\partial}{\partial x_j}\left[\bar{\rho}D\frac{\partial \tilde{\xi}}{\partial x_j} + \bar{\rho}(\widetilde{\xi u_j} - \tilde{\xi}\tilde{u}_j)\right], \tag{3}$$

where $D$ refers to the molecular mass diffusion coefficient. With the unity Lewis number assumption, $D$ is calculated through the thermal conductivity as $D = k/\rho C_p$. For turbulent flames, generally, the molecular diffusion is less important due to the strong turbulent transport. Therefore, the Lewis number was assumed to be unity in this work, which is one of the intrinsic assumptions of the CMC model. The assumption has been relaxed in some previous CMC studies [27, 28], and it has been shown that the results with and without this assumption are similar. Here $k$ is the thermal conductivity, calculated using the Eucken approximation [21], i.e. $k = \mu C_v(1.32 + 1.37R/C_v)$. Here $C_v$ is the heat capacity at constant volume and derived from $C_v = C_p - R$. Here $C_p = \sum_{m=1}^{M} Y_m C_{p,m}$ is the heat capacity at constant pressure, and $C_{p,m}$ is estimated from JANAF polynomials [22]. A gradient-type model is adopted to estimate the sub-grid scalar flux in Eq. (3), i.e.

$$\bar{\rho}(\widetilde{\xi u_j} - \tilde{\xi}\tilde{u}_j) = -\bar{\rho}D_t \frac{\partial \tilde{\xi}}{\partial x_j}. \tag{4}$$

Here $D_t = \mu_t/\bar{\rho}Sc_t$ is sub-grid scale diffusivity with $\mu_t$ being sub-grid scale dynamic viscosity. The $Sc_t$ is turbulent Schmidt number and is assumed as 0.7 [29].

The sub-grid mixture fraction variance $\widetilde{\xi''^2}$ is calculated by an algebraic model [30]

$$\widetilde{\xi''^2} = c_v \Delta^2 \frac{\partial \tilde{\xi}}{\partial x_i} \frac{\partial \tilde{\xi}}{\partial x_i}. \tag{5}$$

$\Delta$ is the LES filter width and is taken as the cube root of the LES cell volume, i.e. $\Delta = V_{LES}^{1/3}$. $V_{LES}$ is the volume of a LES cell. The constant $c_v$ in Eq. (5) is assumed to be 0.1 [30]. Moreover, the scalar dissipation rate $\widetilde{N}$ includes the contributions from the resolved mixture fraction field (i.e. $\widetilde{N}_{res}$) and the sub-grid one (i.e. $\widetilde{N}_{sgs}$) [31]

$$\widetilde{N} = \widetilde{N}_{res} + \widetilde{N}_{sgs} = \underbrace{\frac{\partial \tilde{\xi}}{\partial x_i} \frac{\partial \tilde{\xi}}{\partial x_i} D}_{resolved} + \underbrace{\frac{c_N}{2} \frac{\mu_t}{\bar{\rho} \Delta^2} \widetilde{\xi''^2}}_{sub-grid}. \tag{6}$$

The sub-grid scale scalar dissipation rate $\widetilde{N}_{sgs}$ is modelled based on the assumption that a characteristic velocity timescale is proportional to a characteristic mixing time scale [31-33], which is parameterized by the model constant $c_N$ in Eq. (6). This constant characterizes the contribution of scalar dissipation from the sub-grid field on the total one, i.e. $\widetilde{N}$. Tyliszczak [16] performed a systematic analysis on the sensitivity of CMC modelling of the Cabra flame to this constant [5] and the results suggested that the lift-off height can be accurately reproduced when $c_N = 120$, which is followed in our LES−CMC simulations.

2.2 *Conditional moment closure model*

The governing equation for the conditionally filtered species mass fraction can be written as [33-35]

$$\frac{\partial Q_\alpha}{\partial t} + \widetilde{u_k|\eta} \frac{\partial Q_\alpha}{\partial x_k} = \widetilde{N|\eta} \frac{\partial^2 Q_\alpha}{\partial \eta^2} + \widetilde{\omega_\alpha|\eta} - \frac{1}{\overline{\rho|\eta} \tilde{P}(\eta)} \frac{\partial}{\partial x_k} [\overline{\rho|\eta} \tilde{P}(\eta)(\widetilde{u_k Y_\alpha|\eta} - \widetilde{u_k|\eta} Q_\alpha)], \tag{7}$$

where $Q_\alpha \equiv \widetilde{Y|\eta}$ is the conditional filtered mass fraction of $\alpha$-th species. $\eta$ is the sample space variable for mixture fraction, whereas the operator "$(\cdot|\eta)$" means conditioning on mixture fraction. $\widetilde{u_k|\eta}$, $\widetilde{N|\eta}$ and $\widetilde{\omega_\alpha|\eta}$ are the conditional filtered velocity, scalar dissipation rate and reaction rate of α-th species, respectively. The filtered density function $\widetilde{P}(\eta)$ is assumed to be $\beta$-shaped and calculated with the filtered mixture fraction $\widetilde{\xi}$ and its variance $\widetilde{\xi''^2}$. $\overline{\rho|\eta}$ is conditionally filtered density. Note that the CMC equation is practically solved on a different grid resolution $\Delta_{CMC}$ from the LES one $\Delta_{LES}$ and normally the former is coarser than the latter [35-37]. Therefore, the LES and CMC equations are filtered with various filter sizes.

The second term on the LHS can be divided into two terms

$$\widetilde{u_k|\eta}\frac{\partial Q_\alpha}{\partial x_k} = \frac{\partial}{\partial x_k}\left(\widetilde{u_k|\eta}Q_\alpha\right) - Q_\alpha\frac{\partial \widetilde{u_k|\eta}}{\partial x_k}. \tag{8}$$

The last term on the RHS of Eq. (7) can be modeled with a gradient-type model [37]

$$\widetilde{u_k Y_\alpha|\eta} - \widetilde{u_k|\eta}Q_\alpha = -D_t\frac{\partial Q_\alpha}{\partial x_k}. \tag{9}$$

Neglecting $\overline{\rho|\eta}\widetilde{P}(\eta)$ and substituting Eqs. (8) and (9) into Eq. (7), one can obtain the following governing equation for $Q_\alpha$

$$\frac{\partial Q_\alpha}{\partial t} + \frac{\partial}{\partial x_k}\left(\widetilde{u_k|\eta}Q_\alpha\right) = Q_\alpha\frac{\partial \widetilde{u_k|\eta}}{\partial x_k} + \widetilde{N|\eta}\frac{\partial^2 Q_\alpha}{\partial \eta^2} + \widetilde{\omega_\alpha|\eta} + \frac{\partial}{\partial x_k}\left(D_t\frac{\partial Q_\alpha}{\partial x_k}\right). \tag{10}$$

The product of conditional density and filtered density function in the last term (sub-grid diffusion) in the RHS of Eq. (7) is moved of the spatial derivative and therefore cancelled out, resulting in its form in Eq. (10). This simplification is based on the following reasons: (1) when the value of $\widetilde{P}(\eta)$ equals zero, the last term in Eq. (7) numerically tends to be infinity, and this may increase the calculation uncertainties; (2) the diffusion flux through one CMC cell is actually summed from all the LES faces constituting that CMC cell in our implementation (see Section 2.3), and the LES mesh resolution is sufficiently small to make sense of neglecting the gradient of $\overline{\rho|\eta}\widetilde{P}(\eta)$ across these faces. This has been used in our previous work [18, 23, 24] and reasonable predictions of the reactive statistics in mixture fraction space and physical space are achieved.

Integrating the above governing equation within a control volume $\Omega^{CMC}$ yields

$$\underbrace{\int_{\Omega^{CMC}} \frac{\partial Q_\alpha}{\partial t} d\Omega}_{T_0} + \underbrace{\int_{\Omega^{CMC}} \frac{\partial}{\partial x_k}(\widetilde{u_k|\eta} Q_\alpha) d\Omega}_{T_1} =$$

$$\underbrace{\int_{\Omega^{CMC}} Q_\alpha \frac{\partial \widetilde{u_k|\eta}}{\partial x_k} d\Omega}_{T_2} + \underbrace{\int_{\Omega^{CMC}} \widetilde{N|\eta} \frac{\partial^2 Q_\alpha}{\partial^2 \eta} d\Omega}_{T_3} + \underbrace{\int_{\Omega^{CMC}} \widetilde{\omega_\alpha|\eta} d\Omega}_{T_4} + \underbrace{\int_{\Omega^{CMC}} \frac{\partial}{\partial x_k}(D_t \frac{\partial Q_\alpha}{\partial x_k}) d\Omega}_{T_5}, \quad (11)$$

where $\Omega^{CMC}$ represents the CMC cell. The first term in the LHS, $T_0$, is unsteady term, whilst the terms $T_1$ and $T_2$ denote conditional convection and dilatation, respectively. $T_3$ represents micro-mixing, $T_4$ chemical reaction, and $T_5$ sub-grid scale conditional scalar flux. In our work, the conditionally filtered total enthalpy $Q_h \equiv \widetilde{h|\eta}$ is solved from

$$\int_{\Omega^{CMC}} \frac{\partial Q_h}{\partial t} d\Omega + \int_{\Omega^{CMC}} \frac{\partial}{\partial x_k}(\widetilde{u_k|\eta} Q_h) d\Omega =$$

$$\int_{\Omega^{CMC}} Q_h \frac{\partial \widetilde{u_k|\eta}}{\partial x_k} d\Omega + \int_{\Omega^{CMC}} \widetilde{N|\eta} \frac{\partial^2 Q_h}{\partial^2 \eta} d\Omega + \int_{\Omega^{CMC}} \frac{\partial}{\partial x_k}(D_t \frac{\partial Q_h}{\partial x_k}) d\Omega, \quad (12)$$

which is similar to Eq. (11) without chemical reaction term $T_4$.

The Amplitude Mapping Closure (AMC) model [38] is employed to model $\widetilde{N|\eta}$ in the LES resolution, i.e.

$$\widetilde{N|\eta} = N_0 G(\eta), \quad (13)$$

$$N_0 = \widetilde{N} / \int_0^1 \widetilde{P}(\eta) G(\eta) d\eta, \quad (14)$$

$$G(\eta) = exp(-2[erf^{-1}(2\eta - 1)]^2). \quad (15)$$

Here $G(\eta)$ is a shape function and calculated from the inverse error function $erf^{-1}(x)$. The conditionally filtered scalar dissipation rate in CMC cells (denoted with superscript "CMC") and they are calculated by integrating $\widetilde{N|\eta}$ over all the LES cells within one CMC cell [33]

$$\widetilde{N|\eta}^{CMC} = \frac{\int_{\Omega^{CMC}} \bar{\rho} \widetilde{P}(\eta) \widetilde{N|\eta} d\Omega}{\int_{\Omega^{CMC}} \bar{\rho} \widetilde{P}(\eta) d\Omega}. \quad (16)$$

The mixture fraction and its variance on a CMC cell, $\tilde{\xi}^{CMC}$ and $\widetilde{\xi''^2}^{CMC}$, are given as [33]

$$\tilde{\xi}^{CMC} = \frac{\int_{\Omega^{CMC}} \bar{\rho} \tilde{\xi} d\Omega}{\int_{\Omega^{CMC}} \bar{\rho} d\Omega}, \quad (17)$$

$$\widetilde{\xi''^2}^{CMC} = \frac{\int_{\Omega CMC} \bar{\rho}\widetilde{\xi''^2} d\Omega}{\int_{\Omega CMC} \bar{\rho} d\Omega} + \frac{\int_{\Omega CMC} \bar{\rho}\widetilde{\xi}^2 d\Omega}{\int_{\Omega CMC} \bar{\rho} d\Omega} - \left(\frac{\int_{\Omega CMC} \bar{\rho}\widetilde{\xi} d\Omega}{\int_{\Omega CMC} \bar{\rho} d\Omega}\right)^2. \tag{18}$$

First-order CMC model is used and hence the conditionally filtered reaction rate can be modelled as $\widetilde{\omega_\alpha|\eta} \approx \omega_\alpha(Q_1, \ldots Q_n, Q_T)$. Here $n$ represents the total number of species, and $Q_T \equiv \widetilde{T|\eta}$ is the conditionally filtered temperature. Finally, the unconditionally filtered variables $\tilde{f}$ (e.g. $\bar{\rho}$, $\tilde{T}$ and $\widetilde{Y_\alpha}$) are obtained by integrating the conditional value $\widetilde{f|\eta}$ in mixture fraction space

$$\tilde{f} = \int_0^1 \widetilde{f|\eta} \widetilde{P}(\eta) d\eta, \tag{19}$$

in which $\widetilde{f|\eta}$ is the conditionally filtered scalars (e.g. $1/\widetilde{\rho|\eta}$, $Q_T$ and $Q_\alpha$).

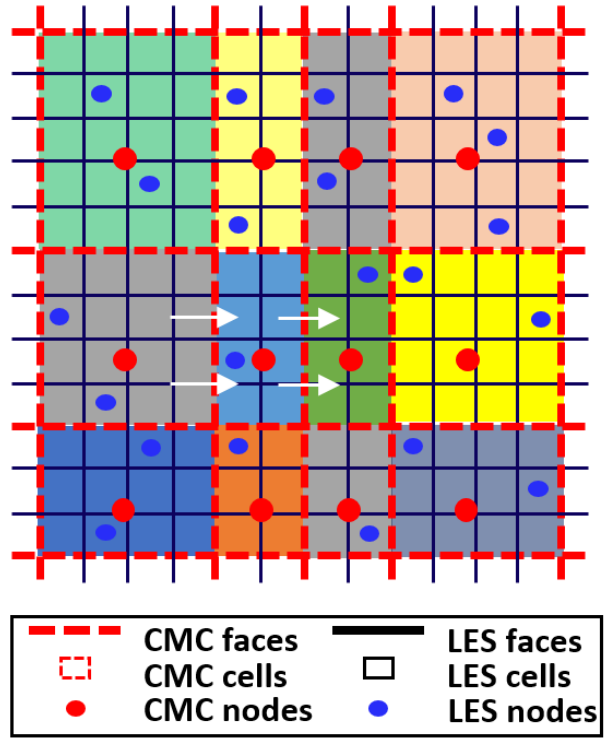

**Fig. 1** Schematic of CMC cell reconstruction from LES mesh. The arrows indicate the directions of convective fluxes of $Q_\alpha$ or $Q_h$. Each LES cell and CMC cell has individual nodes.

*2.3 Finite volume discretization of CMC equations*

For the current LES−CMC formulations, the mesh for CMC equation discretization, $\Omega^{CMC}$, is reconstructed from the LES cells, which is shown in Fig. 1. The CMC nodes (red dots in Fig. 1) are generated within the same domain as that for LES. Then the centroids of the LES cells (i.e. blue dots) search for the host CMC node based on the minimal distance algorithms, i.e. the distance between the centroid of the LES cell and its host CMC node is the smallest. With this, the LES cells have, and only have, unique CMC nodes. Furthermore, the CMC faces (dashed lines in Fig. 1) are selected from the LES faces such that the host CMC nodes of the LES cells sharing them are different. The individual CMC nodes can be enclosed by a sequence of CMC faces. The polyhedral CMC control volumes are therefore generated, and the CMC governing equations, Eqs. (11) and (12), are discretized over them. The finite volume discretization of the individual terms in Eq. (11) are detailed as below.

- Term $T_0$ (unsteadiness)

$$\int_{\Omega^{CMC}} \frac{\partial Q_\alpha}{\partial t} d\Omega \approx \frac{\partial}{\partial t} \int_{\Omega^{CMC}} Q_\alpha \, d\Omega \approx V^{CMC} \frac{\partial Q_\alpha}{\partial t}, \tag{20}$$

where $V^{CMC}$ is the volume of a CMC cell.

- Term $T_1$ (convection)

$$\int_{\Omega^{CMC}} \frac{\partial}{\partial x_k} \left( \widetilde{u_k | \eta} Q_\alpha \right) d\Omega = \oint_{\partial \Omega^{CMC}} \left( \widetilde{u_k | \eta} Q_\alpha \right) \cdot \mathbf{n} dS, \tag{21}$$

$\partial \Omega^{CMC}$ refers to the total faces of a CMC cell.

$$\oint_{\partial \Omega^{CMC}} \left( \widetilde{u_k | \eta} Q_\alpha \right) \cdot \mathbf{n} dS = \sum_{m=1}^{F^{CMC}} \oint_{\partial \Omega^{CMC}} (\widetilde{u_k | \eta} Q_\alpha \cdot \mathbf{n}) \Delta S_m =$$

$$\sum_{m=1}^{F^{CMC}} (\widetilde{u_x | \eta} Q_\alpha n_x + \widetilde{u_y | \eta} Q_\alpha n_y + \widetilde{u_z | \eta} Q_\alpha n_z)_m \Delta S_m, \tag{22}$$

in which $F^{CMC}$ stands for the number of LES faces surrounding the CMC node. The quantities, $n_x$, $n_y$ and $n_z$, are the Cartesian components of the CMC face normal vectors. $\widetilde{u_x|\eta}$, $\widetilde{u_y|\eta}$ and $\widetilde{u_z|\eta}$ are the Cartesian components of $\widetilde{u|\eta}$. The convection fluxes are expected to influence the physical transport between the neighbouring CMC cells, indicated by the arrows as showed in Fig. 1. One can see from Eqs. (21) and (22) that the species flux from one CMC cell would be gained by the neighbouring one and therefore it is conserved.

As mentioned before, $\widetilde{u_k|\eta}$ is modelled as $\tilde{u}_k$ [35], Eq. (22) can be re-written as

$$\sum_{m=1}^{F^{CMC}} \left(\widetilde{u_x|\eta}Q_\alpha n_x + \widetilde{u_y|\eta}Q_\alpha n_y + \widetilde{u_z|\eta}Q_\alpha n_z\right)_m \Delta S_m =$$

$$\sum_{m=1}^{F^{CMC}} (\tilde{u}_x Q_\alpha n_x + \tilde{u}_y Q_\alpha n_y + \tilde{u}_z Q_\alpha n_z)_m \Delta S_m. \qquad (23)$$

The effects of numerical schemes for convection term $T_1$ will be studied in this work, including Upwind (UD) scheme and blended UD / CD (Central Differencing) schemes [39]. Figure 2 shows the face value $\phi_f$ at the face $f$ calculated from its owner and neighbour cells $P$ and $N$:

$$\phi_f = f_x \phi_P + (1 - f_x)\phi_N. \qquad (24)$$

Here, the interpolation factor $f_x$ is defined as the ratio of distances $\overline{fN}$ and $\overline{PN}$:

$$f_x = \frac{\overline{fN}}{\overline{PN}}. \qquad (25)$$

The differencing scheme using Eq. (24) to determine the face value is central differencing.

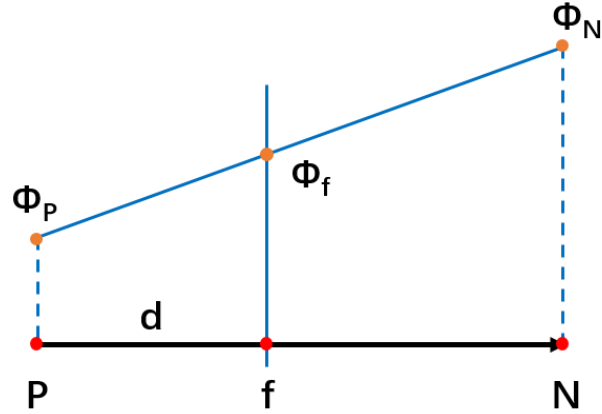

**Fig. 2** Schematic of surface interpolation [40].

For upwind differencing scheme, the face value $\phi_f$ is determined according to the direction of mass flux $F$:

$$\phi_f = \begin{cases} \phi_P & for\ F \geq 0 \\ \phi_N & for\ F < 0 \end{cases}. \tag{26}$$

The blended scheme of CD and UD is implemented with a blending factor $\gamma$

$$\phi_m = \gamma(\phi_m)_{CD} + (1-\gamma)(\phi_m)_{UD}, \tag{27}$$

where $\phi_m$ is a generic variable $\phi$ (e.g. $\widetilde{u_k|\eta}Q_\alpha$) at the $m$-th face. $(\phi_m)_{CD}$ and $(\phi_m)_{UD}$ are the numerical fluxes predicted with CD and UD schemes, respectively. $\gamma$ is a tuneable parameter, and $\gamma = 1$ denotes CD scheme, whereas $\gamma = 0$ denotes UD scheme.

The effects of mesh non-orthogonality on numerical flux calculations are addressed based on OpenFOAM [40, 41]. According to Jasak [40], non-orthogonality correction is applied to calculate the projection of the gradient over the normal face area vector through a given face $f$, usually associated with the diffusive term of a transport equation.

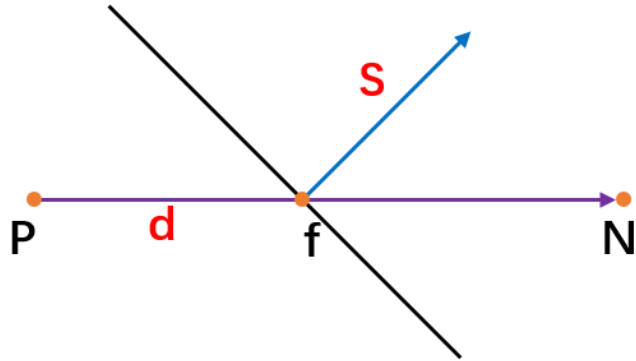

**Fig. 3** Schematic of vectors **d** and **S** on a non-orthogonal mesh.

If the mesh is orthogonal, i.e. vectors **d** and **S** in Fig. 3 are parallel, one can have:

$$\mathbf{S} \cdot (\nabla \phi)_f = |\mathbf{S}| \frac{\phi_N - \phi_P}{|\mathbf{d}|} \tag{28}$$

With it, the face gradient of $\phi$ can be calculated from the two values around the face. For non-orthogonal mesh, in order to make use of the higher accuracy of Eq. (28), the projection of the gradient vector (or tensor) onto the face normal vector $(\mathbf{S} \cdot (\nabla \phi)_f)$ is split into an orthogonal contribution, which is implicitly treated, and a non-orthogonal contribution, which is treated explicitly:

$$\mathbf{S}.(\nabla\phi)_f = \underbrace{\mathbf{\Delta}.(\nabla\phi)_f}_{orthogonal\ contribution} + \underbrace{\mathbf{k}.(\nabla\phi)_f}_{non-orthogonal\ correction} \tag{29}$$

The two vectors introduced in Eq. (29), $\mathbf{\Delta}$ and $\mathbf{k}$, have to satisfy $\mathbf{S} = \mathbf{\Delta} + \mathbf{k}$. The detailed descriptions about the non-orthogonality correction can be found in Ref. [40].

- Term $T_2$ (dilatation)

$$\int_{\Omega^{CMC}} Q_\alpha \frac{\partial \widetilde{u_k|\eta}}{\partial x_k} d\Omega = Q_\alpha \int_{\Omega^{CMC}} \frac{\partial \widetilde{u_k|\eta}}{\partial x_k} d\Omega = Q_\alpha \oint_{\partial\Omega^{CMC}} \widetilde{u_k|\eta} \cdot \mathbf{n} dS, \tag{30}$$

where we assumed that $Q_\alpha$ is constant within a CMC cell [33, 35]. Since $\widetilde{u_k|\eta}$ is modelled as $\tilde{u}_k$ [35], Eq. (30) can be re-written as

$$\oint_{\partial\Omega^{CMC}} \widetilde{u_k|\eta} \cdot \mathbf{n} dS \approx \oint_{\partial\Omega^{CMC}} \tilde{u}_k \cdot \mathbf{n} dS \approx \sum_{m=1}^{F^{CMC}} (\tilde{u}_x n_x + \tilde{u}_y n_y + \tilde{u}_z n_z)_m \Delta S_m, \tag{31}$$

where $\tilde{u}_x$, $\tilde{u}_y$ and $\tilde{u}_z$ are the Cartesian components of filtered velocity $\tilde{u}$ at the $m$-th LES faces, respectively.

- Term $T_3$ (micro-mixing)

$$\int_{\Omega^{CMC}} \widetilde{N|\eta} \frac{\partial^2 Q_\alpha}{\partial^2 \eta} d\Omega \approx V^{CMC} \widetilde{N|\eta} \frac{\partial^2 Q_\alpha}{\partial \eta^2}. \tag{32}$$

Here both $\widetilde{N|\eta}$ and $\frac{\partial^2 Q_\alpha}{\partial \eta^2}$ are assumed to constant in one CMC cell.

- Term $T_4$ (chemistry)

$$\int_{\Omega^{CMC}} \widetilde{\omega_\alpha|\eta} d\Omega \approx V^{CMC} \widetilde{\omega_\alpha|\eta}. \tag{33}$$

It is assumed that the reaction rate $\widetilde{\omega_\alpha|\eta}$ does not change within one CMC cell and therefore can be moved out of the integration over the CMC cell.

- Term $T_5$ (sub-grid scale diffusion)

$$\int_{\Omega^{CMC}} \frac{\partial}{\partial x_k}(D_t \nabla Q_\alpha) d\Omega = \oint_{\partial\Omega^{CMC}} (D_t \nabla Q_\alpha) \cdot \mathbf{n} dS =$$

$$\sum_{m=1}^{F^{CMC}} D_{t,m} (\frac{\partial Q_\alpha}{\partial x} n_x + \frac{\partial Q_\alpha}{\partial y} n_y + \frac{\partial Q_\alpha}{\partial z} n_z)_m \Delta S_m, \tag{34}$$

where $D_{t,m}$ is the sub-grid diffusivity at the $m$-th CMC face. The derivatives, $\partial Q_\alpha/\partial x$, $\partial Q_\alpha/\partial y$, and $\partial Q_\alpha/\partial z$, are the Cartesian components of the gradient of $Q_\alpha$. The discretization of the individual CMC terms in Eq. (12) are the same and therefore not repeated here.

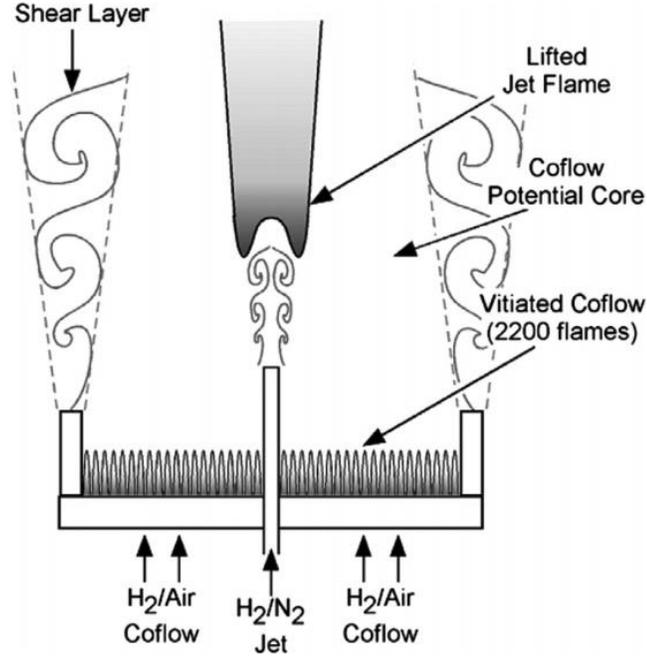

**Fig. 4.** Schematic of the Berkeley $H_2/N_2$ flame [5].

## 3. Flame information and numerical implementation

### 3.1 Berkeley $H_2/N_2$ flame

The lifted $H_2/N_2$ flame in a vitiated co-flowing jet measured by Cabra et al. [5] is simulated in this work. The schematic of this burner is shown in Fig. 4. The central jet of $H_2$ and $N_2$ is injected from a burner with a diameter of $D_j$ = 4.57 mm. A vitiated co-flow of the combustion products from a lean premixed $H_2$/air flame is provided to ignite the central fuel jet. The conditions of the central jet and surrounding co-flow are listed in Table 1. The central jet consists of 25% $H_2$ and 75% $N_2$ (by volume), while the vitiated co-flow is 14.74% $O_2$, 75.34% $N_2$ and 9.89% $H_2O$ (by volume). The temperatures of the fuel and co-flow are 305 K and 1,045 K, respectively. The bulk velocity of the fuel jet is $U_j$ = 107 m/s, while the co-flow velocity is 3.5 m/s. Their Reynold numbers are 23,600 and 18,600, respectively. Furthermore, the stoichiometric mixture fraction $\xi_{st}$ is 0.474, calculated based on Bilger's formulation [42]. There exists a most reactive mixture fraction around which autoignition occurs first due to optimal thermo-chemical conditions [43] and it is about 0.054 for this flame. The measured lift-off height of this flame is about $10D_j$ [5].

Table 1. Boundary conditions at fuel jet and co-flow

| Parameters | Fuel jet | Co-flow |
| --- | --- | --- |
| Diameter | 4.57 mm ($D_j$) | 210 mm |
| Temperature | 305 K | 1,045 K |
| Velocity | 107 m/s ($U_j$) | 3.5 m/s |
| Mole fraction of $H_2$ | 0.25 | 0 |
| Mole fraction of $O_2$ | 0 | 0.1474 |
| Mole fraction of $N_2$ | 0.75 | 0.7534 |
| Mole fraction of $H_2O$ | 0 | 0.0989 |
| Reynold Number | 23,600 | 18,600 |

*3.2 Numerical implementation*

A cylindrical domain is used for LES and CMC simulations. It starts from the burner exit plane, and the domain size in the axial, radial and azimuthal directions are $30D_j \times 10D_j \times 2\pi$, respectively. The coordinate origin lies at the center of the $H_2/N_2$ fuel jet. Different CMC resolutions will be studied in this work, which will be detailed in Section 3.3. Besides, mixture fraction space is discretized by 51 nodes with two boundaries at $\eta = 0$ (co-flow) and $\eta = 1$ (fuel jet), respectively, and the nodes are clustered around the stoichiometric and most reactive mixture fractions.

In the LES, for the fuel jet, one-seventh power law is applied for the mean axial velocity of the fuel jet, consistent with the experiments [5]. The synthetic eddy method [44] is used to reproduce the turbulence and the Reynolds stress components are estimated following Masri et al. [45]. Top-hat profile is given for the co-flow mean velocity. The mixture fraction is unity at the jet, while zero at the co-flow. Zero-gradient condition is enforced for all the quantities at the lateral and outlet boundaries.

For the CMC boundaries in the physical space, mixing solutions of $Q_\alpha$ and $Q_h$ are specified at both fuel jet and co-flow. The thermo-chemical compositions at the two boundaries, i.e. $\eta = 0$ and $\eta = 1$, follow the conditions of fuel jet and co-flow tabulated in Table 1. Zero-gradient condition for $Q_\alpha$ and $Q_h$ is applied at the lateral and outlet boundaries. The CMC cells in the interior domain are initialized by the mixing solutions.

The LES governing equations, i.e. Eqs. (1)–(3), are solved with OpenFOAM®, whilst the CMC equations (Eqs. 11 and 12) are solved by an in-house CMC solver developed at University of Cambridge [18, 19, 22, 24]. The two solvers are interfaced through on-the-fly data exchange (e.g. filtered density and temperature) at each time step, following the strategies detailed in Refs. [18, 19, 22, 24]. The PIMPLE algorithm[†] is adopted for the coupling between velocity and pressure in LES, and first-order implicit Euler scheme is used for time discretization. Both convection and diffusion terms in the LES equations are discretized by central differencing scheme. For the CMC equations, the second-order central differencing is used for sub-grid diffusion term, whereas the linear interpolation is applied for dilatation term. The micro-mixing term in Eqs. (11) and (12) is calculated with TDMA (Tridiagonal Matrix Algorithm) method, and the ODE solver VODPK [46] is used for the chemical reaction terms $\widetilde{\omega_\alpha|\eta}$. Different schemes for convection term in the CMC equations will be studied, and the detailed information is presented in Section 3.3. A chemical mechanism of 10 species and 23 reactions is used for hydrogen oxidation [47], which is also used in Refs. [48, 49] for modelling the same flame. The time step for both LES and CMC solvers is $10^{-6}$ s. 48 bi-processors 2.60 GHz cores are used from ASPIRE 1

---

[†] In OpenFOAM®, PIMPLE algorithm is a combination of PISO (Pressure Implicit with Splitting of Operator) and SIMPLE (Semi-Implicit Method for Pressure-Linked Equations) methods.

Cluster from National Supercomputing Center in Singapore. The Flow-Through Time (FTT) of this flame is $T_j = L_x/U_j \approx 1.3$ ms, where $L_x$ is the streamwise length of the computational domain. The statistical results in Section 4 are collected over 10 FTT after the initial field effects are purged (over 10 FTT).

3.3 *Simulation case*

In this work, the LES mesh is 134×54×42 hexahedral cells and its sufficiency in resolving the flow kinetic energy and conserved scalar mixing can be confirmed by a mesh sensitivity and turbulence length scale estimations in Appendices A-C. To explore the CMC resolution effects, three CMC meshes are studied in Cases 1−3, respectively. Their detailed information is summarized in Table 2. The base (intermediate) CMC mesh in Case 1 contains 94×36×24 cells, whilst the fine CMC mesh in Case 2 has 134×54×42 and the coarse one in Case 3 consists of 54×36×24 cells. They are differentiated with CMC resolutions in the flame region, i.e. $10D_j \times 1.5D_j \times 2\pi$. The approximated ratios of the LES cells to CMC cells in Cases 1−3 are 4, 1 and 8, respectively. Note that in Case 2 the CMC mesh is the same as that of the LES mesh. Meanwhile, three convection schemes are considered, including UD and blended UD / CD schemes. These are compared through Cases 1, 4 and 5, as tabulated in Table 2.

Table 2. Information on the simulation cases

| Case | Numerical Scheme | CMC mesh | LES cell number per CMC cell |
|---|---|---|---|
| 1 (base) | Blended scheme ($\gamma = 0.7$) | 94×36×24 | 4 |
| 2 | Blended scheme ($\gamma = 0.7$) | 134×54×42 | 1 |
| 3 | Blended scheme ($\gamma = 0.7$) | 54×36×24 | 8 |
| 4 | Blended scheme ($\gamma = 0.5$) | 94×36×24 | 4 |
| 5 | Upwind scheme | 94×36×24 | 4 |

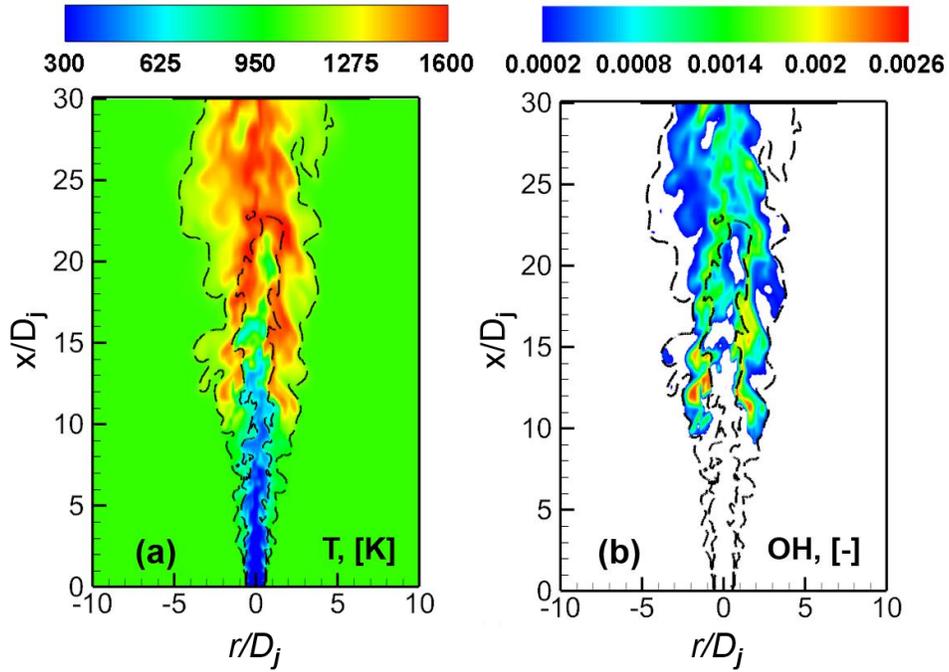

**Fig. 5.** Resolved (a) temperature and (b) OH mass fraction. The inner isolines correspond to the stoichiometric mixture fraction ($\xi_{st} = 0.474$), while the outer isolines the most reactive mixture fraction ($\xi_{mr} = 0.054$).

## 4. Results and discussion

### 4.1 *Basic flame structure*

Figure 5 shows the resolved profiles of temperature and OH mass fraction of Case 1 when the flame stabilizes. It is found that when $x/D_j < 10$, there is no obvious increase of temperature and OH mass fraction, indicating limited chemical reactions there. Nevertheless, when $x/D_j \geq 10$, high temperature and OH mass fractions are observable. This implies that the flame has ignited and been lifted beyond a critical height. Based on Fig. 5, the instantaneous lift-off height is around $10D_j$. This is determined as the minimal axial distance where the OH mass fraction reaches $2\times10^{-4}$ [50, 51]. Almost the same lift-off heights are obtained if other criteria (e.g. $\tilde{T} > 1800$ K) are used. Our results also indicate that the flame base fluctuates between $8.7D_j$ and $11.5D_j$, similar to the experimental observations [52]. Similar fluctuations ($3-4D_j$) of the lift-off height are also reported by the previous LES−CMC simulations [14]. A closer inspection of Fig. 5 reveals that the OH radical is largely presented closer to the isolines of $\xi_{mr}$ at the base. This indicates that the flame is stabilized with local auto-ignition around $10D_j$.

The time-averaged temperature and OH mass fraction from Case 1 are plotted in Fig. 6. It is seen that the mean lift-off height is about $10D_j$, which is consistent with the measurement in the experiment [5]. Note that this height is predicted without tuning the co-flow temperature or velocity, considering their possible uncertainties [5, 6]. In the previous LES of the same flame with finite-differencing CMC model, the pronounced deviations (with the errors of $\pm 5D_j$) are observed for the lift-off height [13, 14]. The mean temperature rises near the fuel-lean mixture conditions, which is also seen from the resolved temperature in Fig. 5(a). The mean OH mass fraction increases since the lift-off height. It peaks between $12D_j$ and $17D_j$ and at the fuel-lean side of the stoichiometric mixture fraction isolines.

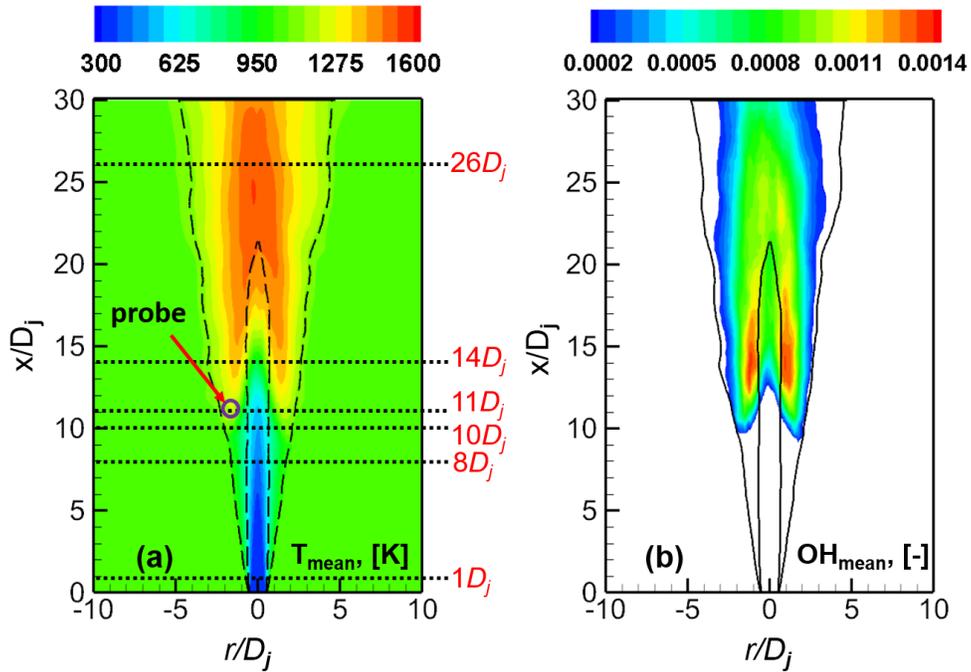

**Fig. 6.** Time-averaged (a) temperature and (b) OH mass fraction. Legend for iso-lines same as in Fig. 5. Probe coordinate: $x = 11D_j$, $y = 1.4D_j$, $z = 0$.

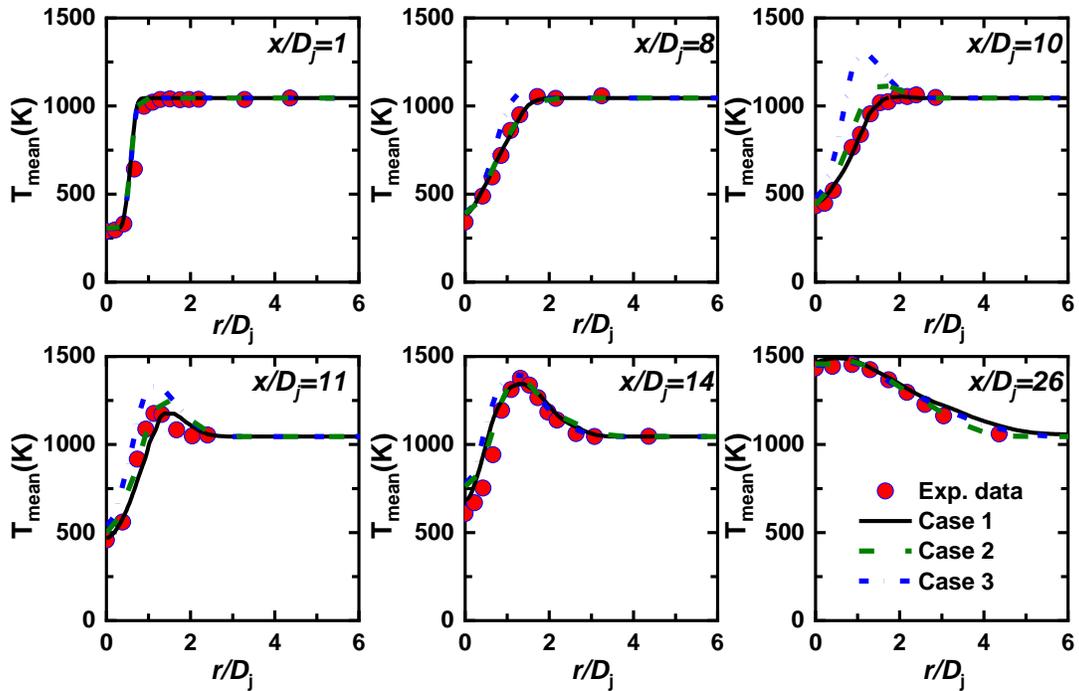

**Fig. 7.** Radial profiles of time-averaged temperature at six axial locations. Experimental data from Ref. [5].

## 4.2 *Effects of CMC resolution*

Three CMC meshes (i.e. Cases 1−3) are used to study the CMC resolution effects. Cases 1−3 respectively correspond to intermediate (base), fine and coarse resolutions, as listed in Table 2. Figure 7 shows the radial profiles of mean temperature at six locations in Cases 1, 2 and 3. At $x/D_j$ = 1, 14 and 26, the radial profiles of the mean temperature in Cases 1, 2 and 3 are in good accordance with the experimental data. However, the mean temperature from coarse CMC mesh, i.e. Case 3, are evidently over-predicted at the axial locations $x/D_j$ = 8, 10 and 11. For instance, the mean temperature between $r/D_j$ < 2 at $x/D_j$ = 10 is higher than the experimental data, indicating that the chemical reactions have been initiated there. This implies that the flame base stabilizes more upstream when the coarse CMC mesh is used, which results in lower lift-off height (about 7.7$D_j$) in Case 3. Additionally, the radial profiles of the mean temperatures predicted with medium and fine CMC meshes (i.e. Cases 1 and 2) almost have no differences at $x/D_j$ = 8, 10 and 11. This corroborates the accuracy in predict the unsteady flame dynamics when non-consistent LES and CMC resolutions are used [33, 35].

Figure 8 shows the radial distributions of the temperature RMS at four locations, i.e. $x/D_j$ = 8, 10, 11 and 14. Here the RMS values are calculated based on the resolved temperature. It is seen that the trends of the radial profiles of the temperature RMS at four axial locations are well captured with in Cases 1−3. The predictions of the coarse CMC mesh in Case 3 show largest deviations from the measurements than the other two at $x/D_j$ = 8 and 10. However, at $x/D_j$ = 11 and 14, the differences of the temperature RMS are relatively small, although the computed temperature RMS slightly deviates from the measured data.

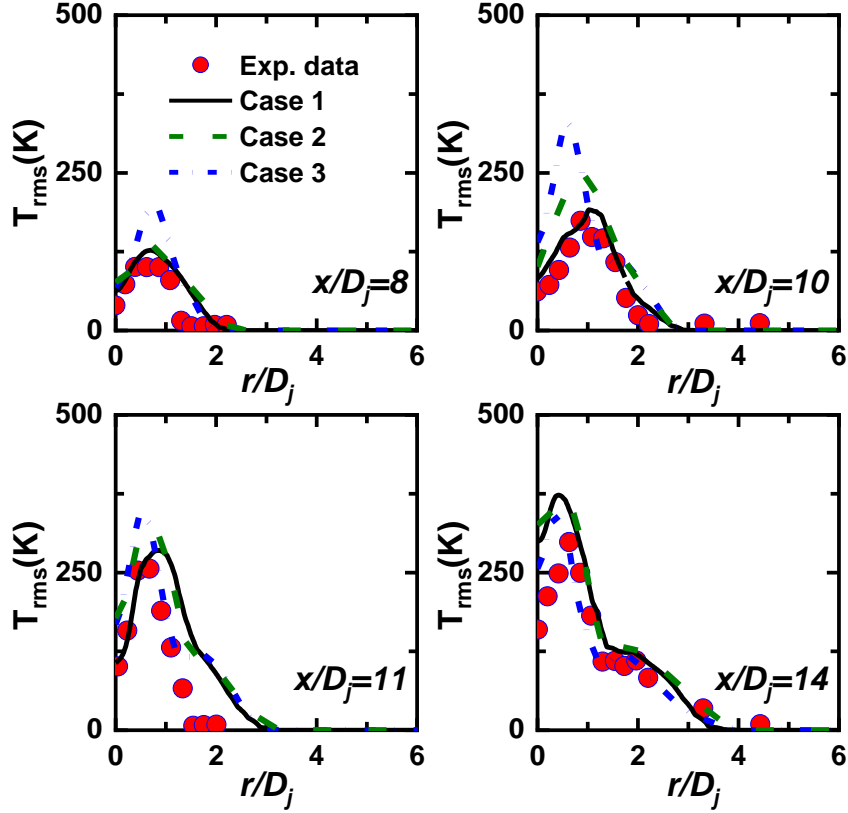

**Fig. 8.** Radial profiles of temperature RMS at four axial locations. Experimental data from Ref. [5].

The radial profiles of the statistics of $H_2$ and OH mass fractions at four axial locations are presented in Figs. 9 and 10, respectively. In Fig. 9, the mean $H_2$ mass fractions in Cases 1 and 2 are basically consistent with the measurements, while hydrogen consumption in Case 3 is slightly overestimated at $x/D_j = 10$, which is associated with the lower lift-off height predicted with the coarse mesh. The radial distributions of the RMS values of $H_2$ mass fraction show some deviations from the measurements, particularly at $x/D_j = 10$ and 11. The OH radical is an important indicator for the beginning of the hydrogen auto-ignition [7]. The predictions with the coarse CMC mesh suggest that the flame is initiated at around $7.7D_j$. Furthermore, the predicted mean and RMS of the OH mass fractions from the coarse CMC mesh are noticeably larger than the measurements at $x/D_j = 8$, 10 and 11. Differences of the mean OH mass fractions between the measurements and Cases 1 and 2 are small compared with those in the coarse mesh.

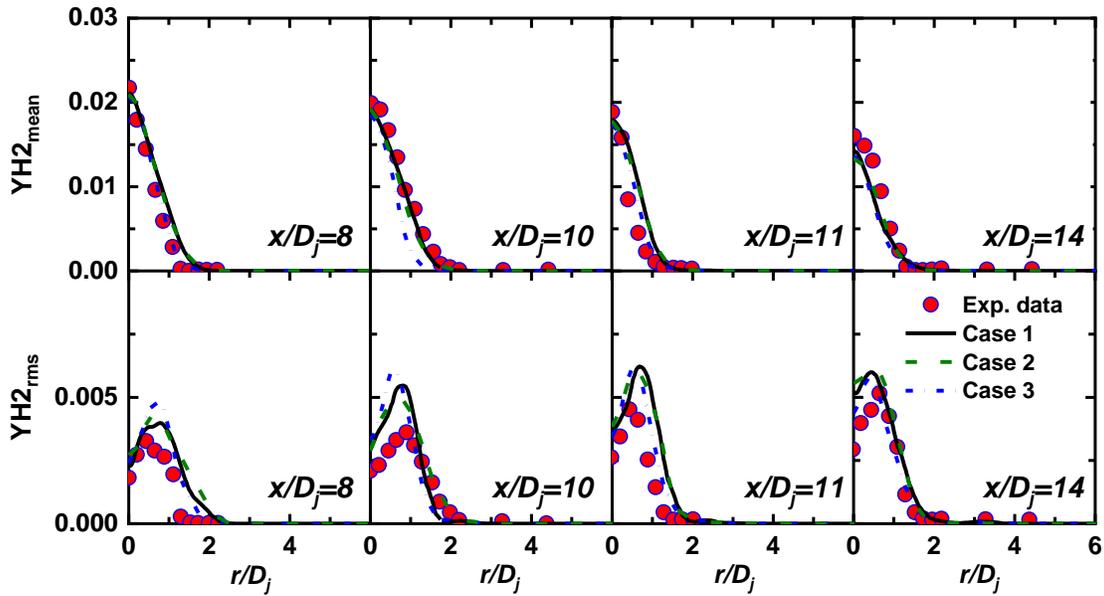

**Fig. 9.** Radial profiles of mean (first row) and RMS (second row) of $H_2$ mass fraction at four axial locations. Experimental data from Ref. [5].

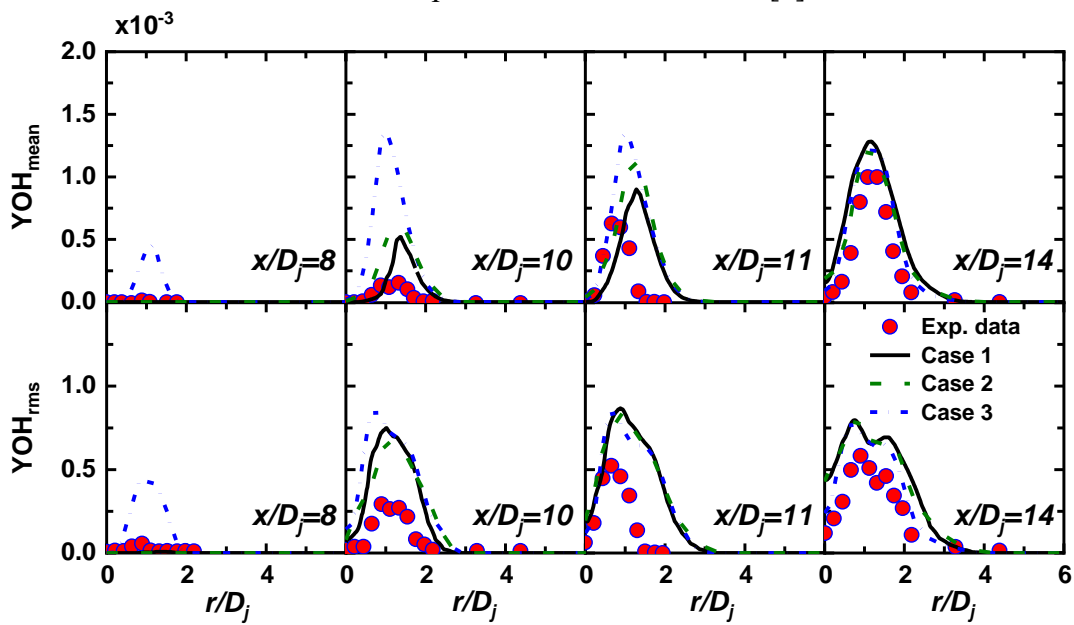

**Fig. 10.** Radial profiles of mean (first row) and RMS (second row) of OH mass fraction at four axial locations. Experimental data from Ref. [5].

The comparison between the axial profiles of temperature statistics predicted from LES-MMC [53]

and this work (Case 1) are illustrated in Fig. 11. It can be found that the current LES-CMC model has the comparable accuracies with the LES-MMC model. The computational cost of the current LES-CMC simulation is slightly more expensive than that with the sparse-Lagrangian MMC method [53] (private communication, M. Cleary), but good parallelization efficiency of our CMC solver significantly reduces the computational cost, which makes the LES simulations affordable.

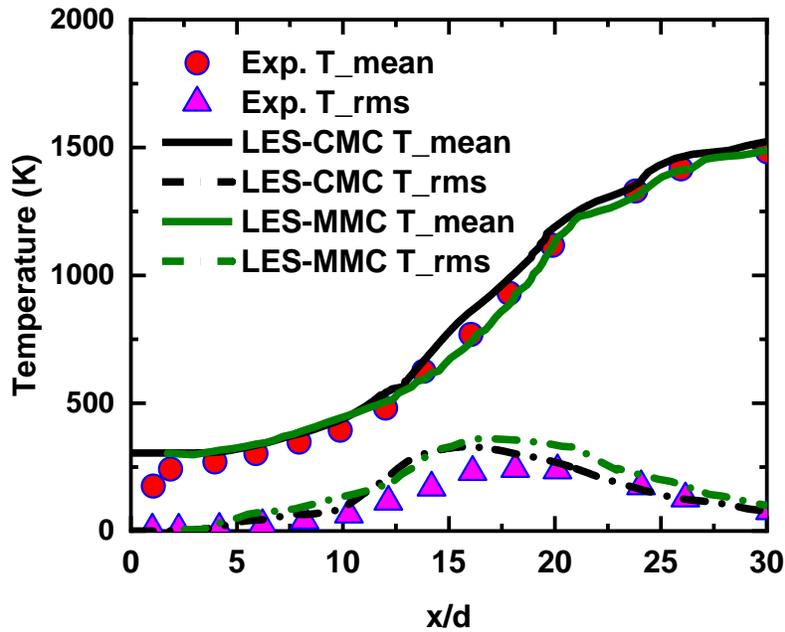

**Fig. 11**. Comparison between the axial temperature statistics predicted with LES-MMC [53] and LES-CMC models.

The influences of CMC mesh resolution on conditionally filtered mass fraction are shown in Fig. 12. Here conditionally filtered OH mass fractions at stoichiometry ($\widetilde{Y_{OH}|\xi}_{st}$) is selected, as OH is an important indicator of auto-ignition event. Demonstration of $\widetilde{Y_{OH}|\xi}_{st}$ in the LES mesh is equivalent to visualizing that in the CMC resolution, since each LES cell in its host CMC cell has the same solutions. It can be found from Fig. 12 that the instantaneous distributions of $\widetilde{Y_{OH}|\xi}_{st}$ around the flame base look similar in Cases 1 and 2, while that from Case 3 is obviously different. Specifically, the flame base in

Case 3 is lower than those in the other two, and meanwhile the gradient of $\widetilde{Y_{OH}|\xi}_{st}$ at the flame base is smaller those in Cases 1 and 2.

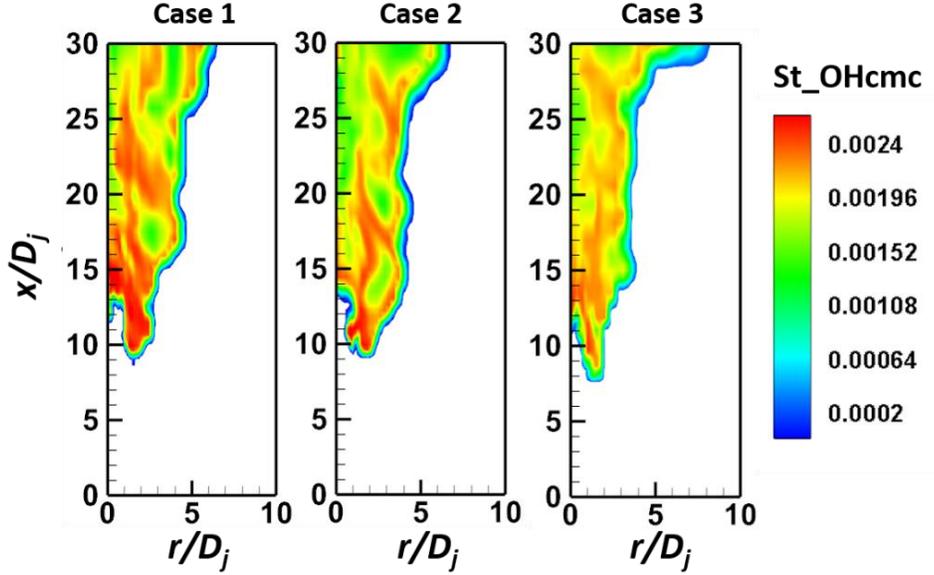

**Fig. 12** Contours of instantaneous conditionally filtered OH mass fractions at stoichiometry from different CMC meshes.

To quantitatively analyze the effects of the CMC mesh resolution on the conditional flame structures, the scatter data from the experiments [5] and the conditional mean temperature at three axial locations ($x/D_j$ = 8, 10 and 14) are illustrated in Fig. 13. The scatter data, including temperature and OH mass fraction, are collected from $r/D_j$ = 0.066 to 1.88 along the radial direction at the abovementioned axial locations, and the simulation results are also extracted from the same locations. At $x/D_j$ = 8, the conditional mean temperature in Cases 1 and 2 shows good accordance with the measurements, while those in Case 3 are over-predicted. At $x/D_j$ = 10, near the flame base, differences between the measured and calculated conditional mean temperature in Cases 1 and 2 are small. The temperature scatters at $x/D_j$ = 10 show that the mixing solutions are dominant at this location, although there are some points away from the mixing line. The latter corresponds to the possible instantaneous localized extinction and re-ignition near the flame base. Over-predictions of the conditional mean temperature in Case 3 at the three locations indicate that higher reactivity (earlier autoignition) at these locations due to the coarse CMC mesh used.

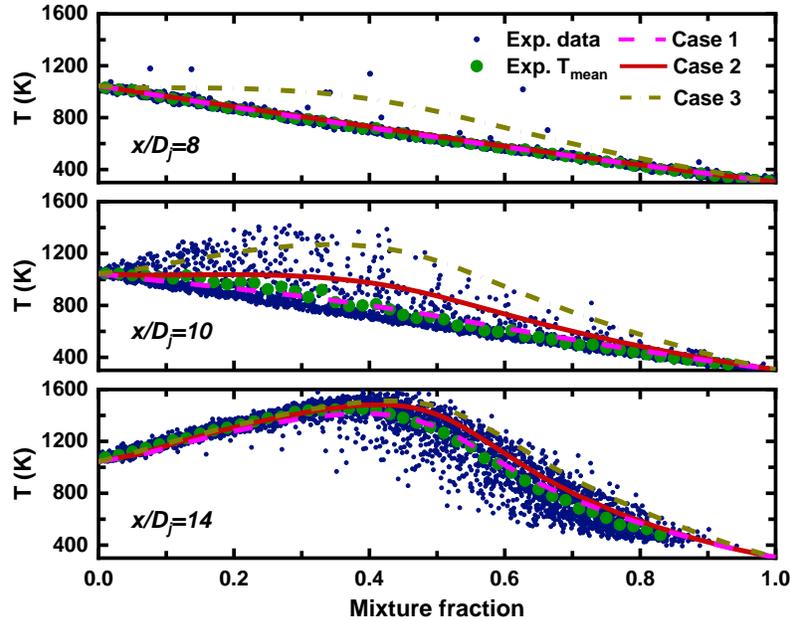

**Fig. 13.** Scatter and conditional mean of temperature at three axial locations. Experimental data from Ref. [5].

Figure 14 shows the temperature fluctuations with respect to the conditional mean values in Cases 1−3 at the same axial locations as in Fig. 13. It can be found that the predicted temperature fluctuations of Case 2 are consistent with the measured values. This may be because the same LES and CMC mesh is used, which removes the need for data transfer between two meshes and leads to high resolution for conditional reactive scalars (e.g. temperature). Moreover, the results from Case 1 are closer to the experimental data than those from Case 3 at the three locations. Generally, increasing CMC mesh resolution enhances the accuracy of the LES−CMC model in capturing the unsteady temperature evolutions.

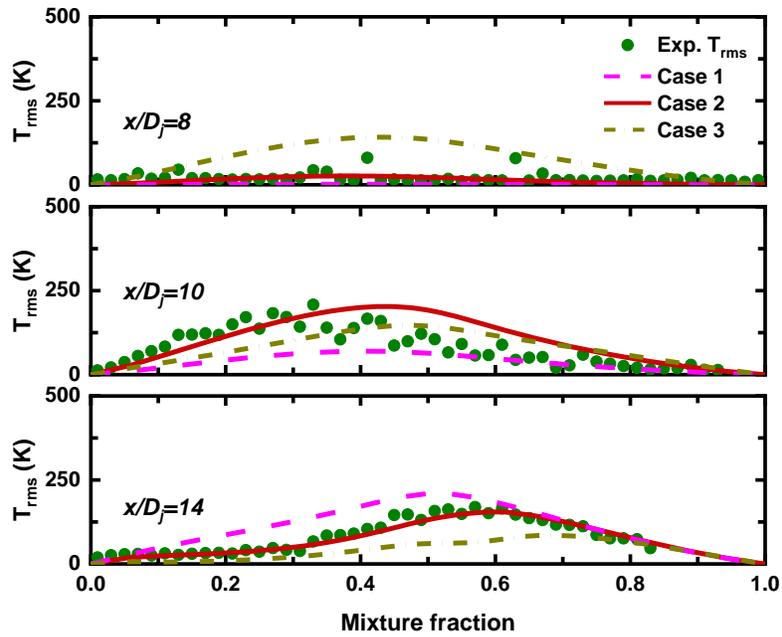

**Fig. 14.** Conditional temperature fluctuation at three axial locations. Experimental data from Ref. [5].

The conditional means and fluctuations of OH mass fraction at the same three locations are demonstrated in Figs. 15 and 16, respectively. It is apparent from Figs. 15 and 16 that the conditional mean OH mass fractions are sensitive to the CMC mesh resolution. The deviations of the OH mass fractions between the measurements and the predictions of Cases 1 and 2 are small compared with those of Case 3, similar to the tendency in Fig. 13. This is consistent with the comparison of the unconditional OH mass fraction in Fig. 10. Similar to the temperature fluctuations in Fig. 14, the fluctuations of conditionally mean OH mass fraction of Case 2 show better agreements with the measurements comparing with Cases 1 and 3.

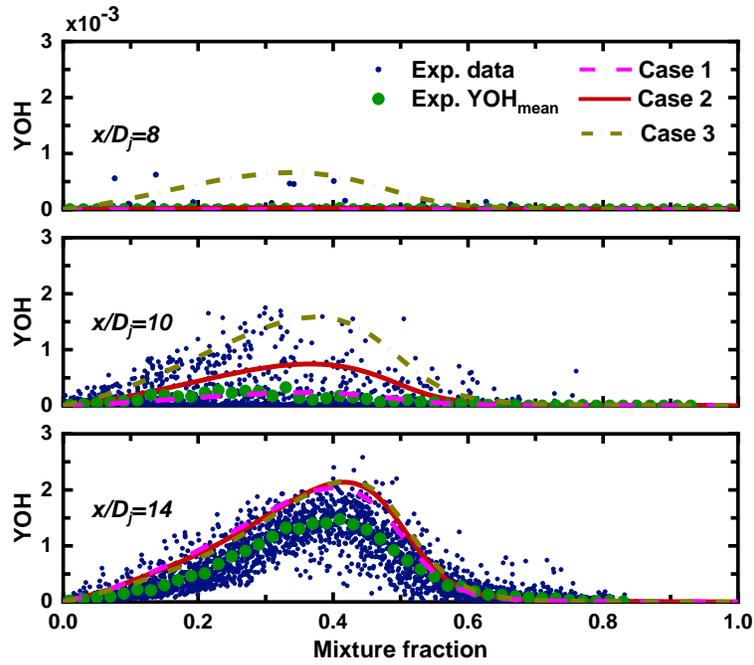

**Fig. 15.** Conditional mean OH mass fraction at three axial locations. Experimental data from Ref. [5].

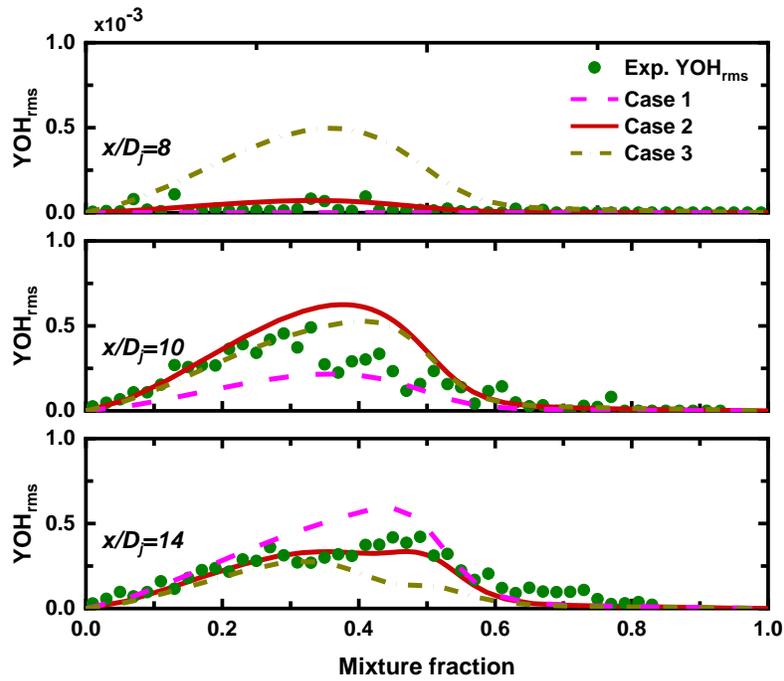

**Fig. 16.** Conditional fluctuation of OH mass fraction at three axial locations. Experimental data from Ref. [5].

Figure 17 shows the time evolutions of OH mass fractions at most reactive ($\xi_{mr}$) and stoichiometric

($\xi_{st}$) mixture fractions. They are extracted from the probed CMC cell ($x/D_j = 11$, $y/D_j = 1.4$, $z/D_j = 0$, marked in Fig. 6a) in Cases 1−3. It can be found that this cell in Case 3 first achieves fully burning state at around 0.0058 s, while in Cases 1 and 2 ignition in the same location is initiated around 0.0085 s and 0.008 s, respectively. The phenomenon is in accordance with the flame solution in physical space as shown in Fig. 10, attributed to the fact that the flame lift-off height calculated from the coarse CMC mesh (Case 3) is lower than that from the refiner meshes (Cases 1 and 2). Meanwhile, more remarkable fluctuations of OH mass fractions at both mixture fractions are captured in Cases 1 and 2 compared with Case 3, as the finer CMC mesh can capture more details of turbulence. It is also seen from Cases 1 and 2 that the variations of the most reactive OH mass fraction are basically consistent with the variations of the stoichiometric OH mass fraction. However, this tendency is not clear in Case 3, although some small variations are also present.

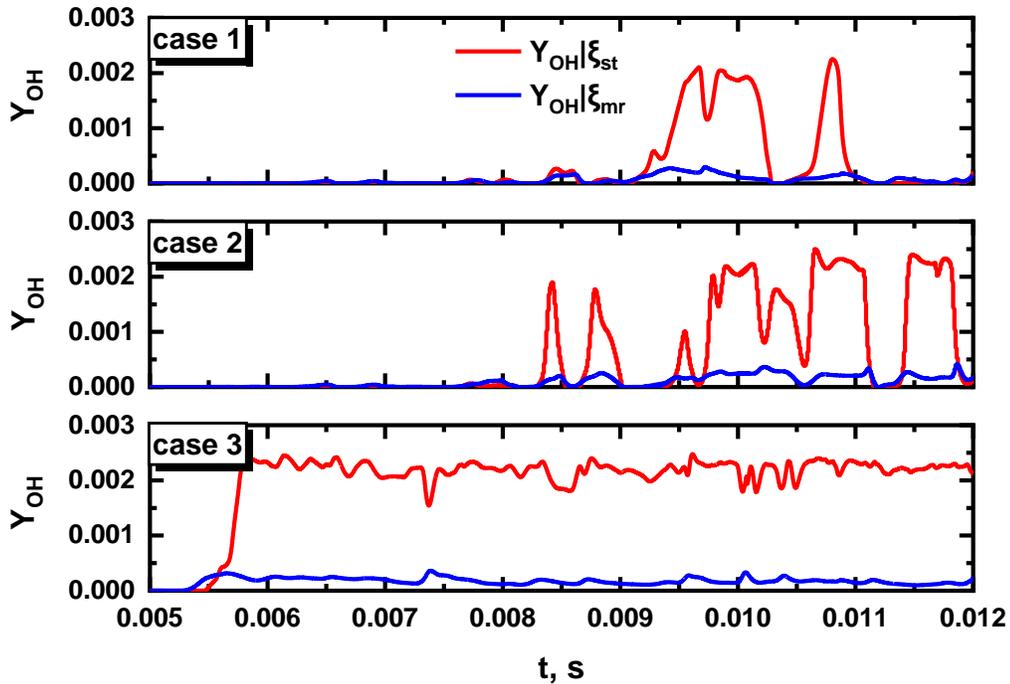

Fig. 17. Time history of the most reactive and the stoichiometric OH mass fraction. Results from the probe ($x/D_j = 11$, $y/D_j = 1.4$, $z/D_j = 0$) shown in Fig. 4(a).

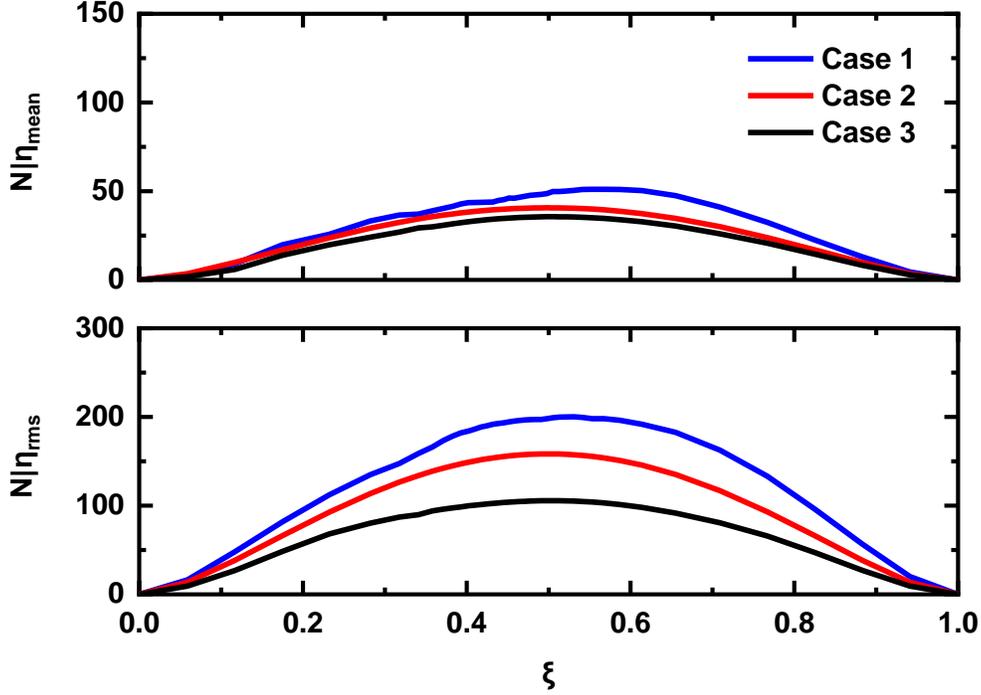

**Fig. 18.** Mean and RMS of conditional scalar dissipation rate ($\widetilde{N|\eta}^{CMC}$) in mixture fraction space. Results from the probe ($x/D_j = 11$, $y/D_j = 1.4$, $z/D_j = 0$) shown in Fig. 4(a).

The distributions of the mean and RMS conditional scalar dissipation rate in mixture fraction space at the same probe for Cases 1−3 are showed in Fig. 18. They are collected from the CMC resolution (i.e. $\widetilde{N|\eta}^{CMC}$) and correspond to Eq. (16). The data were averaged from $t = 0.005$ s to $0.012$ s. Obviously, the mean and RMS of scalar dissipation rate ($\widetilde{N|\eta}^{CMC}$) in the whole mixture fraction range are the largest in Case 1, while they are the smallest in Case 3. The reader should be reminded that in Case 2 the identical LES and CMC meshes are used, and hence the data averaging from LES to CMC mesh (i.e. Eq. 16) is actually not enforced. Nevertheless, the coarse CMC mesh leads to the deviations of $\widetilde{N|\eta}^{CMC}$ relative to that from Case 2. Relatively low or intermediate scalar dissipation rate indicates a high propensity for the mixture to be ignited, as discussed by Mastorakos [43]. This trend has also been demonstrated in Fig. 17. It is also found due to the various mesh resolutions, the differences in $N|\eta_{mean}$ are smaller than differences in $N|\eta_{rms}$, and the difference in mean values from the Cases 1, 2 and 3 is less obvious. Besides, the finest mesh does not lead to a smallest conditionally averaged dissipation rate ($\widetilde{N|\eta}$), and the reason could be attributed to the calculations of $\widetilde{N|\eta}$. As described in Sections 2.1 and 2.2, the $\widetilde{N|\eta}$ in one CMC cell is integrated from its host LES cells. However, the numbers of LES cells in one CMC cell among the three cases are not the same. Moreover, since the probe is close to the flame base, strong unsteadiness and spatial variations of the conditional quantities may exist.

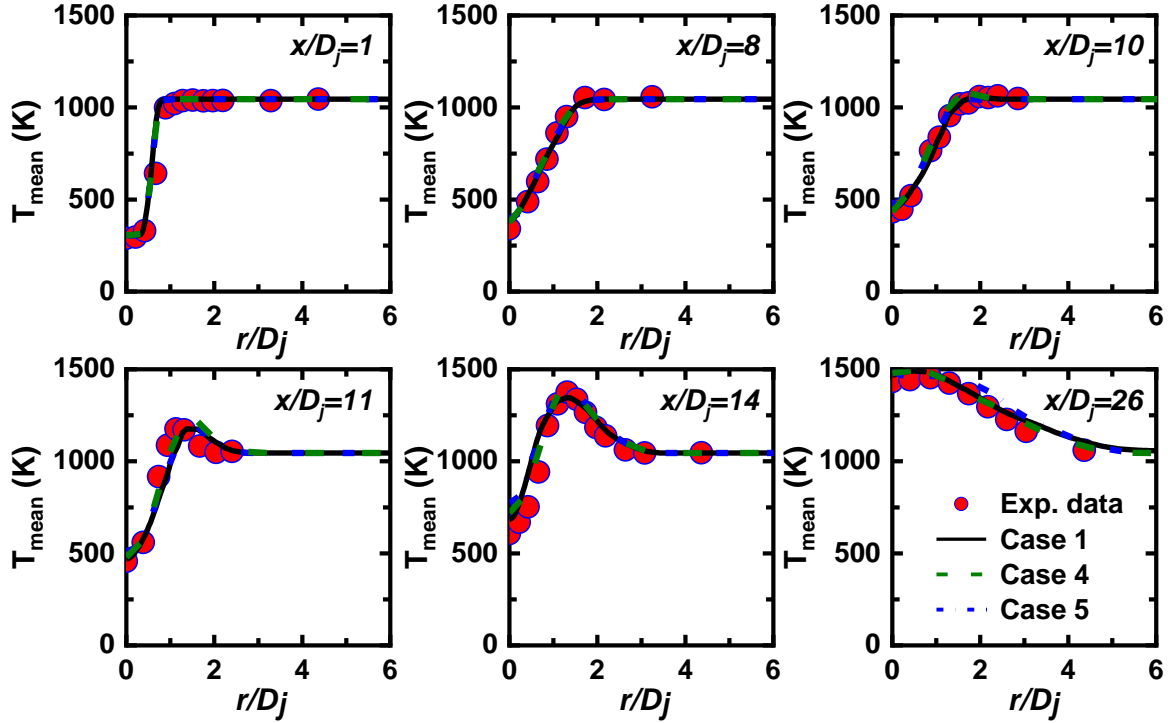

**Fig. 19.** Radial profiles of mean temperature with various convection schemes. Experimental data from Ref. [5].

4.3 *Effects of numerical scheme for CMC equations*

Figure 19 shows the radial profiles of the mean temperature at different axial locations. Cases 1, 4 and 5 respectively correspond to three various convection schemes for the CMC equations, i.e. blended UD / CD schemes with blending factor $\gamma$ = 0.7 and 0.5, as well as UD scheme. In general, there are limited differences between the measured and the simulated mean temperature with three convection schemes. An in-depth observation on Fig. 19 shows that the mean temperatures predicted with blend UD

/ CD schemes are smaller than those calculated with UD scheme at $x/D_j$ = 26. This is because the UD scheme is more dissipative than the blend ones, especially at further downstream locations (e.g. $x/D_j$ = 26) where the mesh size is larger. The radial distributions of temperature RMS values at four locations are presented in Fig. 20. The variations of the temperature RMS values are all reproduced satisfactorily with the three schemes, except a slight overestimation at $x/D_j$ = 11 near the flame base.

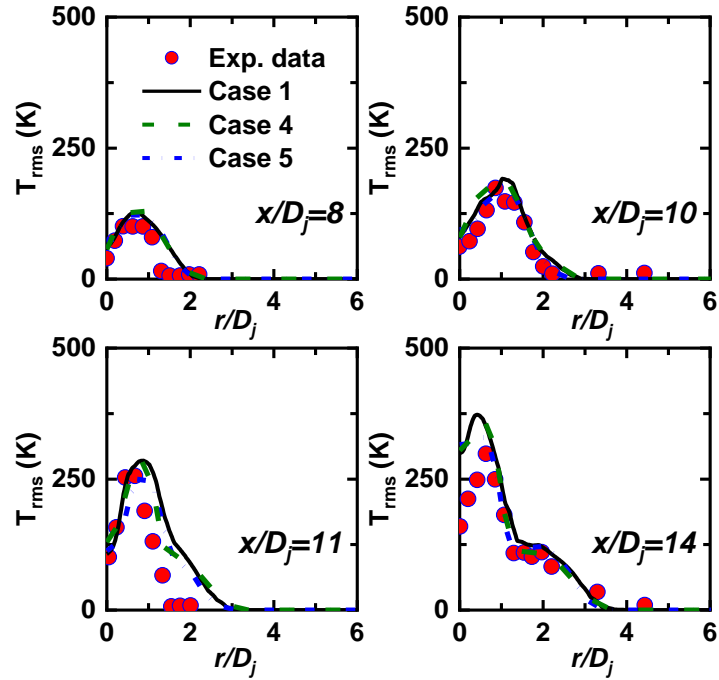

**Fig. 20.** Radial profiles of temperature RMS with different convection schemes. Experimental data from Ref. [5].

The mean and RMS of H$_2$ mass fraction computed with three convection schemes in Cases 1, 4 and 5 are in good agreement with the experimental data, as showed in Fig. 21. The predictions of the RMS of H$_2$ mass fraction show slight difference for the three cases, especially at $x/D_j$ = 10 and 11 where the flame base locates. The fluctuations of H$_2$ mass fraction in Case 1 have the largest deviation from the measurements. Figure 22 shows that the deviations of OH mass fraction between the simulations and measurements gradually increase from $x/D_j$ = 10 and then decrease at the downstream of the flow ($x/D_j$

= 14). This can be ascribed to the slightly overestimated lift-off heights calculated in Case 1, 4 and 5. Similar to the predictions of $H_2$ mass fraction, the fluctuations of OH mass fraction also show a larger deviation from the measurements in Case 1 compared with Cases 4 and 5.

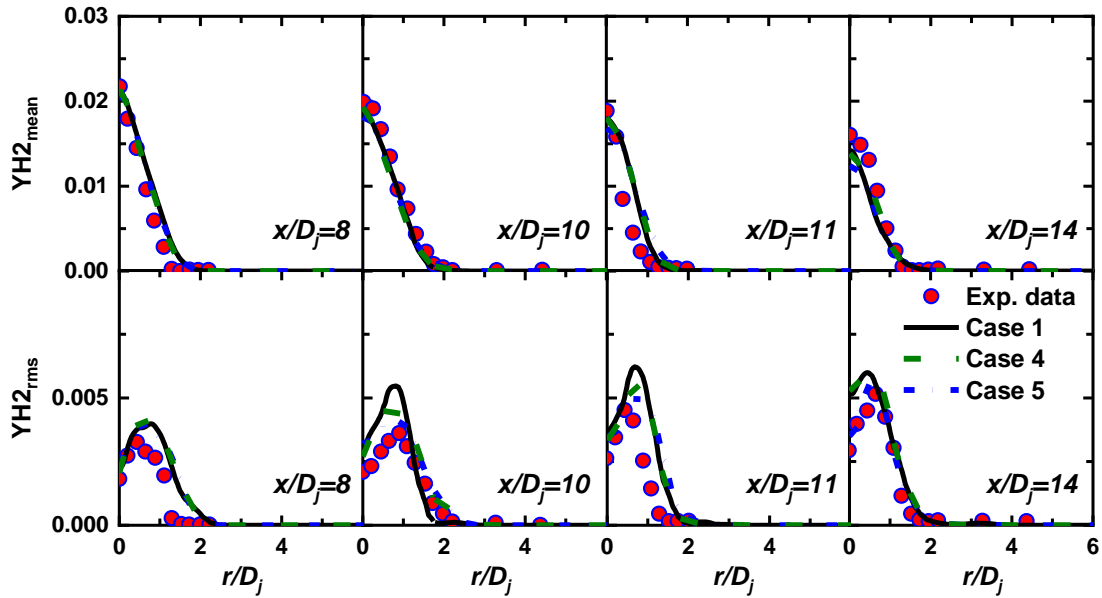

**Fig. 21.** Radial profiles of mean and RMS of $H_2$ mass fraction in Cases 1, 4 and 5 at four axial locations. Experimental data from Ref. [5].

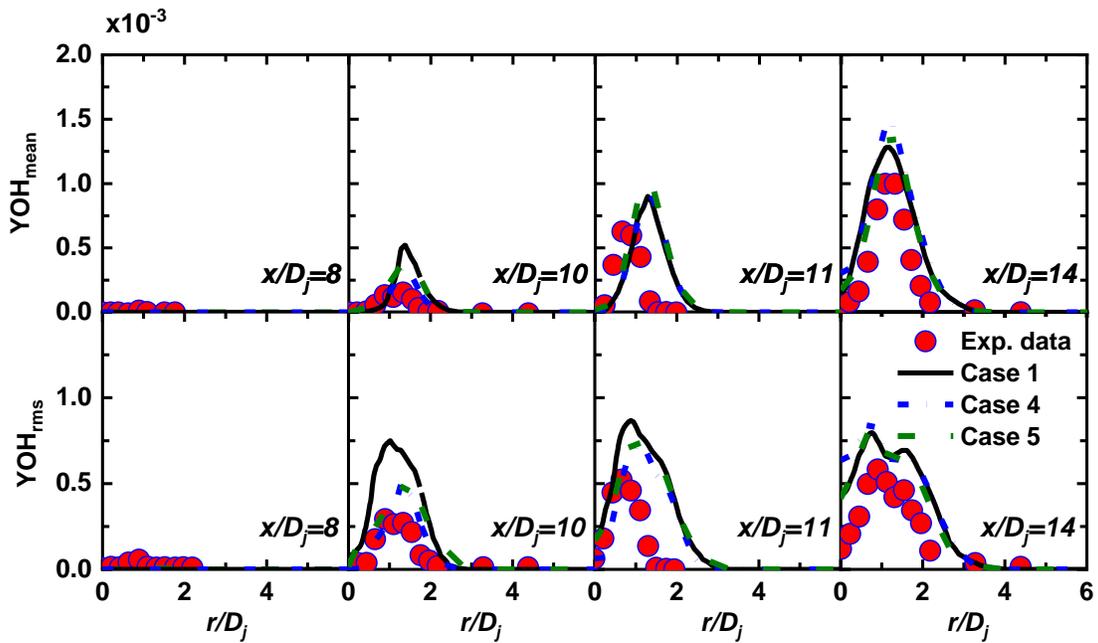

**Fig. 22.** Radial profiles of mean and RMS of OH mass fraction in Cases 1, 4 and 5 at four axial locations. Experimental data from Ref. [5].

The interactions between the CMC cells may considerably affect the predictions of the flame dynamics [14, 21]. As reported in Refs. [14, 16], flame autoignition mechanism can be clarified by the instantaneous magnitudes of different terms in the CMC equations. Therefore, the budgets of CMC terms ($T_1$, $T_2$, $T_3$, $T_4$ and $T_5$) in mixture fraction space will be analyzed in one CMC cell ($x/D_j = 11$, $y/D_j = 1.4$, $z/D_j = 0$, marked in Fig. 6a) near the flame base. We select six time instants during the auto-ignition process with the time interval of 0.00005 s, and the first instant is termed as $t_0$ as shown in Figs. 23, 24 and 25. The contributions of the CMC terms on the conditionally filtered OH mass fraction ($\widetilde{Y_{OH}|\eta}$) in Case 1 (blend factor $\gamma = 0.7$) are shown in Fig. 23, where $t_0$ corresponds to the time of 0.0099s. It is observed that the dilatation, micro-mixing and chemistry terms, i.e. $T_2 = Q_\alpha \frac{\partial \widetilde{u_k|\eta}}{\partial x_k}$, $T_3 = \widetilde{N|\eta} \frac{\partial^2 Q_\alpha}{\partial \eta^2}$ and $T_4 = \widetilde{\omega_\alpha|\eta}$, are close to zero at $t = t_0$, while the convection ($T_1 = \frac{\partial(\widetilde{u_k|\eta} Q_\alpha)}{\partial x_k}$) and sub-grid diffusion terms ($T_5 = \frac{\partial}{\partial x_k}(D_t \frac{\partial Q_\alpha}{\partial x_k})$) have similar magnitudes at $t_0$. Since the $T_1$ and $T_5$ are respectively at the left and right sides of Eq. (11), production of OH radical is negligible at $t_0$. The contribution of the chemistry term merges from $t = t_0 + 50\Delta t$ (Fig. 23b), and progressively increases with time until the flame becomes stable after $200\Delta t$. The micro-mixing term exhibits its contribution after $100\Delta t$, and shows an opposite effect on $\widetilde{Y_{OH}|\eta}$ compared with the chemistry term ($T_4$), particularly at $t = t_0 + 250\Delta t$. The above phenomenon is also reported in Ref. [16]. Both the sub-grid scale diffusion and convection terms have finite effects in Figs. 23(a)−23(d). The convection and diffusion in physical space contribute to the interactions between fresh CMC cell and burning one. However, these contributions are not comparable with that of chemistry term when the maximum temperature in mixture fraction space arrives 1,200 K as showed in Figs. 23(c) and 23(d).

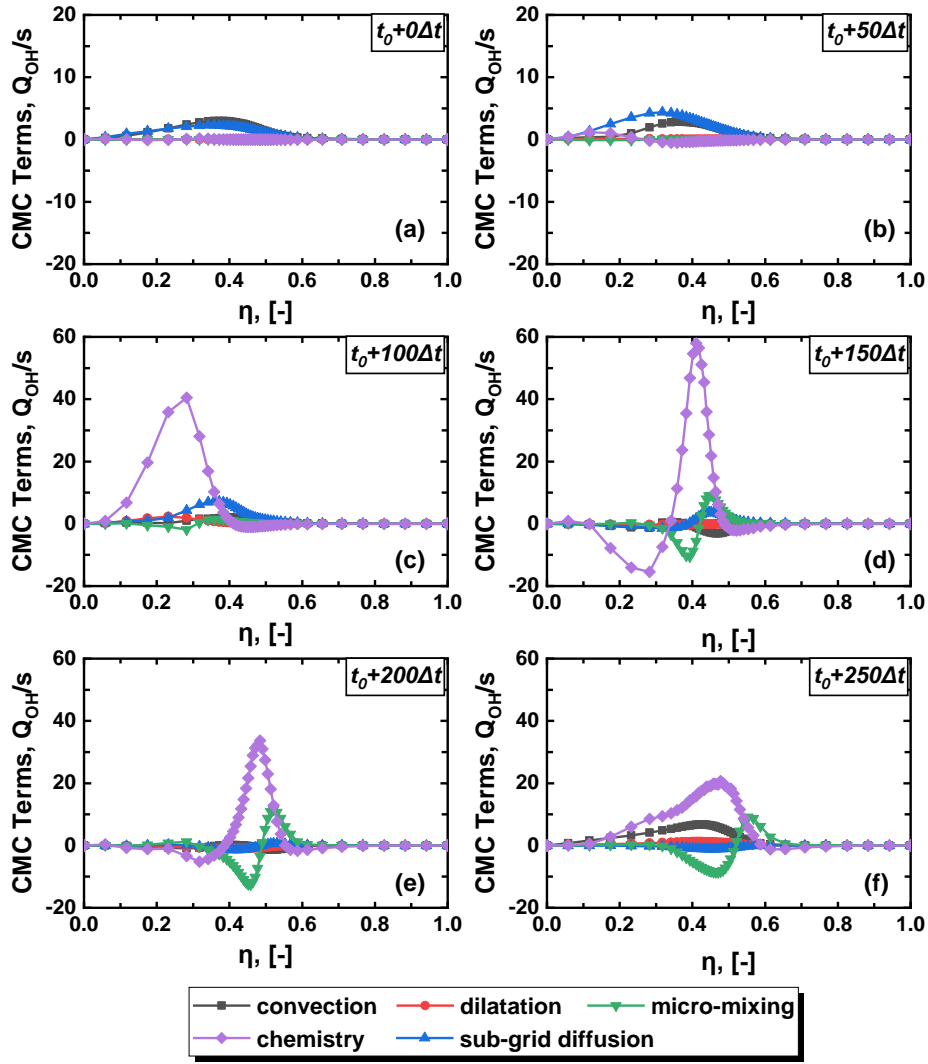

**Fig. 23.** Budgets of the CMC terms for OH mass fraction equation at the probe ($x/D_j = 11$, $y/D_j = 1.4$, $z/D_j = 0$) for Case 1.

Figure 24 shows the balance between different CMC terms for Case 4, where the blended UD / CD scheme with the blending factor of 0.5 is used. The time 0.0134 s is selected as the initial time $t_0$. The contributions of chemistry and micro-mixing terms are similar to that in Case 1 (shown in Fig. 23). The dilatation and sub-grid scale diffusion terms are relatively small, which are basically similar to the counterparts in Case 1. However, the convection term is different from that in Case 1 where the blending factor is 0.7. It is seen that the contributions of the convection term are always negative during the whole ignition process, and the effects of convection showed in Figs. 24(c)−24(f) become more evident than those in Figs. 23(c)−23(f). This indicates that the convection schemes with different blending factors influence the contributions of the convective transport. Additionally, the maximum magnitudes of the contribution of the convection term is about 13 observed in Fig. 24(c), which is larger than that in Case 1.

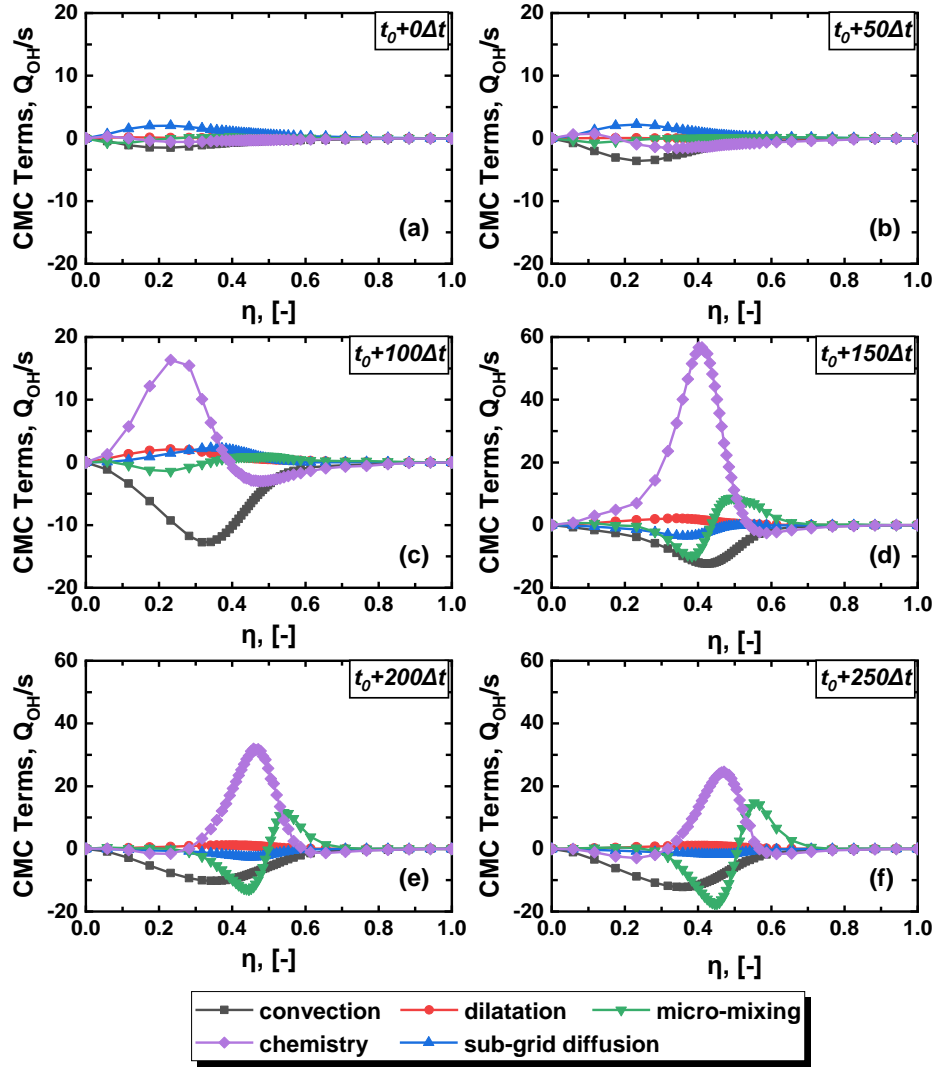

**Fig. 24.** Budgets of the CMC terms for OH mass fraction equation at the probe ($x/D_j = 11$, $y/D_j = 1.4$, $z/D_j = 0$) for Case 4.

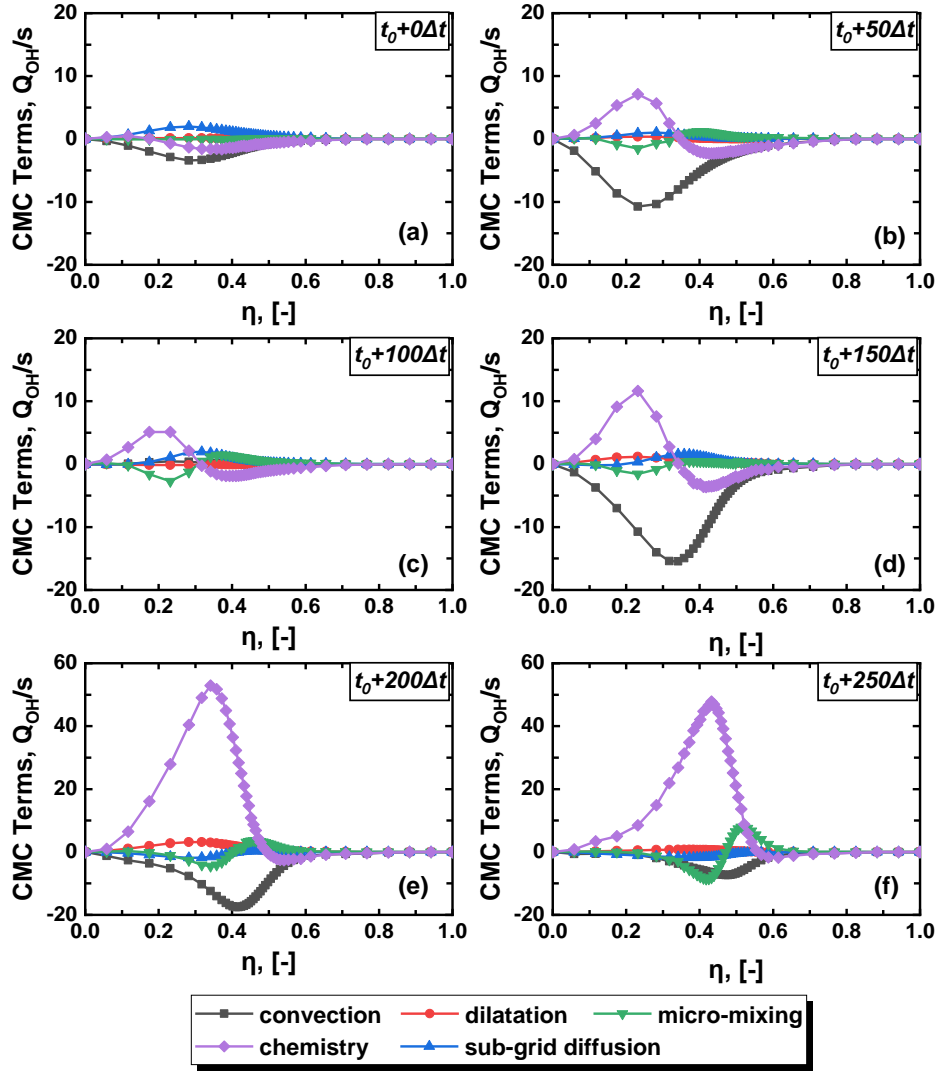

**Fig. 25.** Budgets of the CMC terms for OH mass fraction equation at the probe ($x/D_j = 11$, $y/D_j = 1.4$, $z/D_j = 0$) for Case 5.

Likewise, the budget analysis also conducted for Case 5 in Fig. 25, in which upwind scheme is used. The initial time is $t_0 = 0.00995$ s. Similar to the budget analysis for Cases 4 and 5, the effects of dilatation and sub-grid scale diffusion terms in Case 5 are small during the whole auto-ignition process, while the chemistry term plays an important role on the flame development. The contributions of the micro-mixing term in Fig. 25 are always small and not sufficient to compete with the chemistry term, which is different from Cases 1 and 4. In Cases 1 and 4, the micro-mixing term almost has the same magnitude with the chemistry term when arriving the fully burning state. For the convection term, the absolute value of the maximum contribution of the convection term is about 20 observed in Fig. 25(e), the largest one among those in Cases 1, 4 and 5.

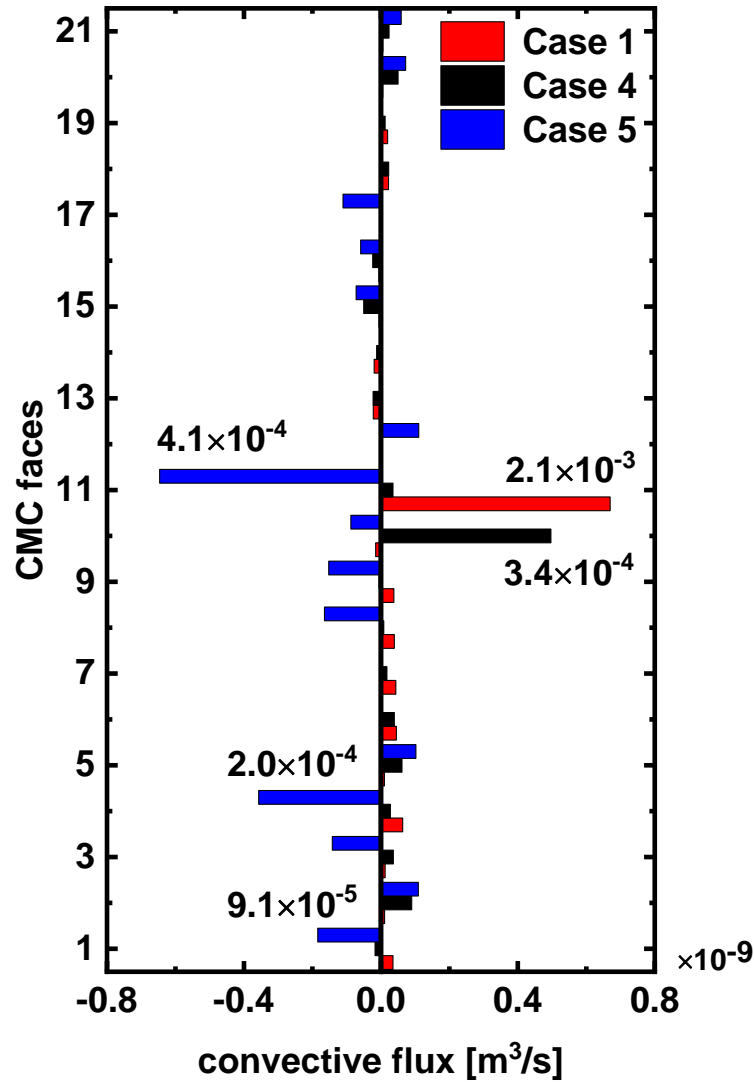

**Fig. 26.** Convective flux across each face of the CMC cell at the probe ($x/D_j = 11$, $y/D_j = 1.4$, $z/D_j = 0$) for Cases 1, 4 and 5. Numbers near the bars are to the OH mass fraction at the stoichiometric mixture fraction ($Y_{OH}|\xi_{st}$) for the neighboring CMC cells.

As mentioned in Section 2.3, one CMC cell contains numerous CMC faces through which the numerical flux enters or leaves the enclosed CMC cell. This would considerably affect the local flame structures and may induce strong unsteadiness of conditional flame structure. The instantaneous convective flux of stoichiometric OH mass fraction from individual CMC faces are illustrated in Fig. 26. Results are extracted from the same probed CMC cell at $t = t_0+50\Delta t$ in Figs. 23(b), 24(b) and 25(b). The negative sign in Fig. 26 means influx of the numerical flux for the CMC cell, while the positive one means the outflux. This CMC cell has 21 CMC faces, and each face is shared by two neighboring CMC cells. The convective flux of blended scheme (Cases 1 and 4) at the CMC faces have a similar trend: the convective flux at most CMC faces are small, and only one face plays a dominant role in convection transport between the neighboring cells. While for the UD scheme (Case 5), most convective fluxes are negative, and there are about half of the CMC faces showing finite and comparable contributions. This further indicates that the reactivity of a CMC cell is more easily to be affected by its neighbors when UD scheme is used.

Besides, OH mass fractions at the stoichiometric mixture fraction ($\widetilde{Y_{OH}|\xi_{st}}$) for the neighboring CMC cells (which share one CMC face with the current CMC cell) are also showed as the numbers near the bars in Fig. 26. For Case 5, the stoichiometric mass fraction $\widetilde{Y_{OH}|\xi_{st}}$ of the current CMC cell is $1.1\times10^{-4}$, while $\widetilde{Y_{OH}|\xi_{st}}$ at its neighbor CMC cells are $4.1\times10^{-4}$ for face 11, $2.0\times10^{-4}$ for face 4 and $9.1\times10^{-5}$ for face 1, respectively. These finite values of $\widetilde{Y_{OH}|\xi_{st}}$ mean that the neighboring CMC cells are under burning state. Therefore, the gross convective fluxes may make the current CMC cell have the propensity to be ignited. For Case 1, $\widetilde{Y_{OH}|\xi_{st}}$ of the current CMC cell is $1.4\times10^{-5}$, while the largest $\widetilde{Y_{OH}|\xi_{st}}$ at its neighbor CMC cell is $2.1\times10^{-3}$ for face 11. Hence the convective fluxes would not instantaneously facilitate the ignition of the current cell. Similar to Case 1, the instantaneous fluxes in Case 4 also do negative contributions to autoignition in the current cell, as the $\widetilde{Y_{OH}|\xi_{st}}$ is $4.5\times10^{-5}$ at this cell, while $Y_{OH}|\xi_{st}$ at its neighbor CMC cell is $3.4\times10^{-4}$ for face 10. However, in Case 1 and 4, this CMC cell still proceeds towards fully burning conditions due to the continuous interactions between the various flame structures in physical space, as illustrated in Figs. 23 and 24.

# 5. Conclusion

The LES−CMC simulations are performed for a lifted $H_2/N_2$ jet in turbulent vitiated co-flow with detailed chemical mechanism. The effects of mesh resolution and numerical scheme of the finite volume CMC model on predictions of reactive scalars and unsteady flame behaviors are analyzed in this work.

The comparisons between the measured and predicted radial distributions of temperature, mixture fraction and species mass fractions at different locations show that the LES−CMC approach has a better performance with the finer CMC mesh. Besides, the lift-off height is underestimated if the CMC mesh resolution is large. However, excessive refinement of the CMC mesh is unable to further improve the prediction accuracy. The time sequences of the most reactive and stoichiometric OH mass fractions in different CMC meshes illustrates that the finer CMC mesh is capable of capturing more unsteady details than the coarser CMC mesh. The coarser CMC mesh has lower conditional scalar dissipation rate, which promotes the ignition of the lifted flame.

Furthermore, the effects of convection schemes for the finite volume CMC equations on the reactive scalar profiles in physical and mixture fraction space are also investigated. Comparisons against the experimental data indicate that the influences of various convection schemes on the time-averaged temperature, mixture fraction and species mass fraction are negligible. The fluctuations of $H_2$ and OH mass fractions show larger deviations from the measurements with hybrid upwind and central differencing scheme, especially around the flame base. The budget analysis illustrates the magnitudes of the individual CMC terms (i.e. $T_1$-$T_5$ in Eq. 11) at various instants. The results suggest that the contributions of the convection term increase with the blending factor, and the absolute value of the maximum contributions of convection term is obtained with the upwind scheme. It is also shown that the instantaneous convective flux of stoichiometric OH mass fraction from individual CMC faces is considerably affected by the convection scheme. The reactivity of a CMC cell is more easily to be affected by its neighbors with upwind scheme.

**Acknowledgement**

This work used the ASPIRE 1 Cluster from National Supercomputing Center, Singapore (https://www.nscc.sg/). GL is sponsored by The CSC Scholarship (201806020055). HZ is supported by Singapore Ministry of Education Tier 1 grant (R-265-000-653-114). Dr. Bertrand Naud from CIEMAT is acknowledged for sharing the experimental data and post-processing routines. The authors thank the anonymous reviewers for the helpful comments.

**Appendix A.** *Sensitivity of reactive scalar statistics to LES mesh resolution*

Figure A1 shows the radial distributions of the time-averaged temperature predicted with two LES meshes (i.e. 260×90×72 and 134×54×42) at six axial locations. Note that the second mesh is used in the above analysis and for brevity they are termed as M1 and M2, respectively. Here the CMC mesh consists of 94×36×24 cells. It is found that the mean temperature from two meshes have good agreements with the measurements at all the locations, although the time-averaged temperature is slightly over-predicted by the fine LES mesh (i.e. M1) in the jet centerline at $x/D_j = 14$. The temperature RMS from the two meshes are compared against the measurements at four positions in Fig. A2. The two results are in good accordance with the experimental data at $x/D_j = 8$ and 10, while the temperature RMS at $x/D_j = 11$ and 14 are overestimated in both M1 and M2. In general, the results from two LES meshes have negligible differences regarding the temperature statistics. Similar tendencies can also be found from the species mass fraction statistics.

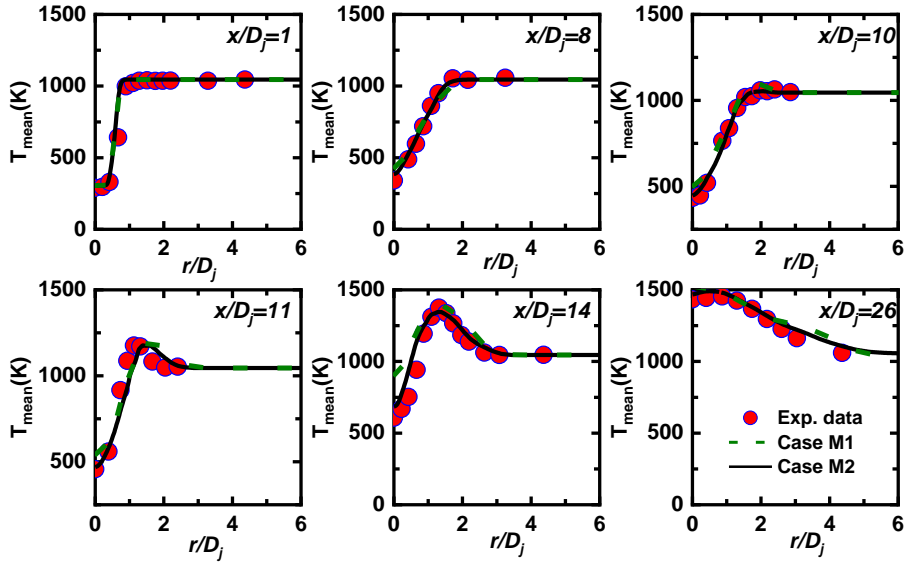

**Fig. A1.** Radial profiles of the time-averaged temperatures from two LES meshes.

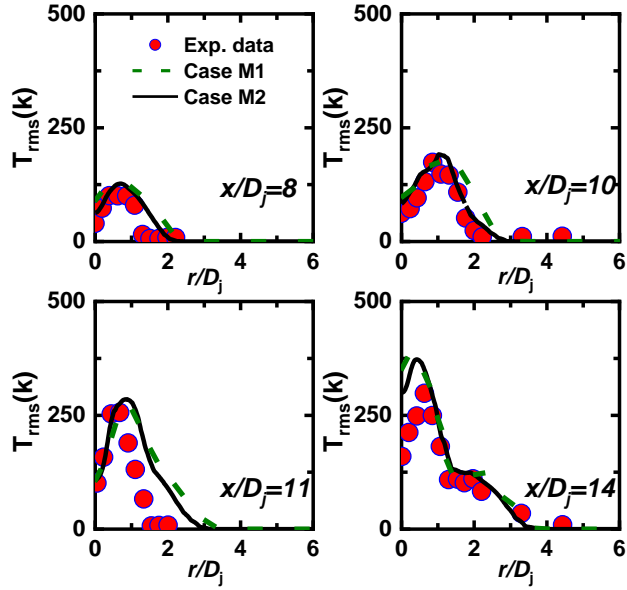

**Fig. A2.** Radial profiles of the temperature RMS from two LES meshes.

Figure A3 shows the mean and RMS of mixture fraction at $x/D_j$ = 8, 10, 11, and 14 from two LES meshes. The distributions of the mean mixture fraction are predicted generally well in both LES meshes at all the locations. Furthermore, the mixture fractions along the centerline at $x/D_j$ = 14 are underestimated with M1. This may be related to the over-prediction of mean temperature at the same location in Fig. A1. Moreover, the RMS of mixture fraction at all the shown locations are well reproduced by the two meshes, in spite of a slight overestimation at $x/D_j = 11$. One can therefore conclude from Figs. A1−A3 that the results from both LES meshes have good agreements with the experimental data, and the mesh (i.e. M2) used in the foregoing studies are sufficient.

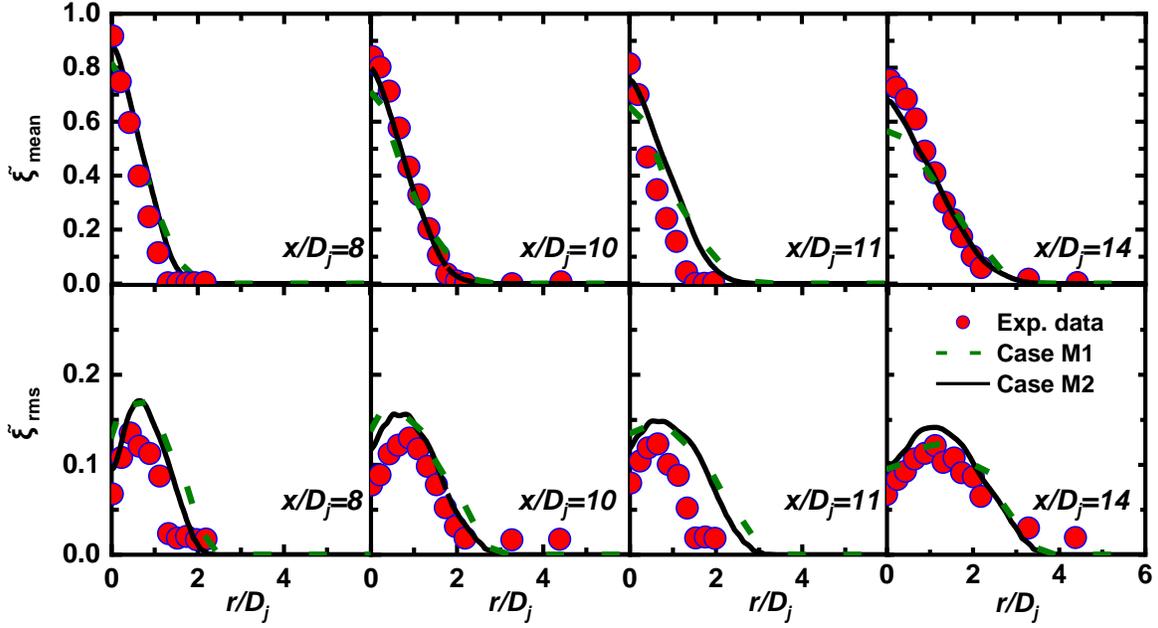

**Fig. A3.** Radial profiles of mean (top) and RMS (bottom) of mixture fraction from two LES meshes.

## Appendix B. *LES mesh resolution compared with turbulent length scales*

The integral length scale ($L_t$) and Kolmogorov scale ($\eta_k$) can be estimated as [54]:

$$L_t \approx D_j(1 + B \cdot x) \tag{A1}$$

$$\eta_k = L_t Re_t^{-3/4} \tag{A2}$$

where $D_j$ is the diameter of a round jet, $B \approx 0.09$ is the expansion rate of the jet [54] and $x$ is the axial coordinate. $Re_t = u'L_t/v$ is the local turbulent Reynolds number [55] obtained using the RMS field of velocity $u'$ from the resolved field and the kinematic viscosity $v$.

Based on Eqs. (A1) and (A2), we compare the nominal LES cell size $\Delta$ with the integral length scale $L_t$ and Kolmogorov scale $\eta_k$ from Case 1 (see Table 2), which are shown in Fig. A4. It is observed that for $x < 20D_j$, the grid size represents less than 8% of the integral length scale and about 10-20 times greater than the Kolmogorov length scale. Since other cases in Table 2 use the same LES resolution, and hence all our simulations are supposed to be sufficient to capture the turbulence scales of large eddies.

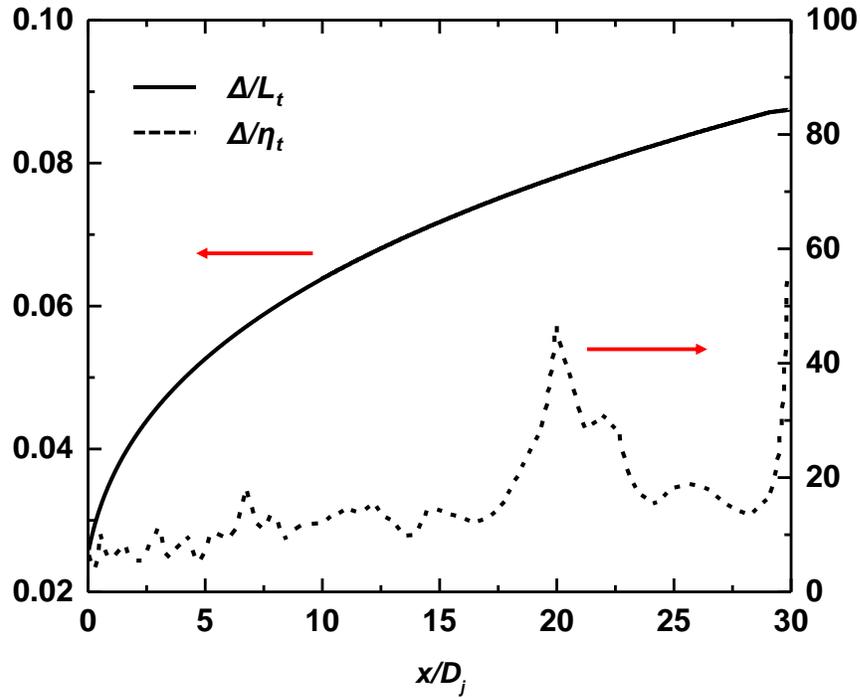

Fig. A4. Nominal LES mesh size ($\Delta$) compared to the integral scale ($L_t$) and the Kolmogorov scale ($\eta_k$) on the jet axis.

## Appendix C. *Fraction of unresolved turbulent kinetic energy*

According to Pope [56], at least 80 % of the total turbulent kinetic energy should be resolved in an LES. The fraction of unresolved kinetic energy is estimated from

$$Me = \frac{k_{sgs}}{k_{RES}+k_{sgs}}, \quad (A3)$$

where $k_{sgs}$ is the sub-grid turbulent kinetic energy and $k_{RES}$ is the resolved turbulent kinetic energy. Figure A5 shows the distribution of $Me$ in different LES meshes, using the data with hydrogen mass fraction $\tilde{Y}_{H2} > 0.001$. The coarse and fine meshes have 260×90×72 and 134×54×42 cells, respectively. The same CMC mesh is used, which consists of 94×36×24 cells. It can be found that the probability density distributions with $Me < 0.2$ are about 0.86 and 0.96 from the coarse and fine meshes, respectively. The results in Fig. A5 further confirm the sufficiency of the LES mesh resolution used in this study.

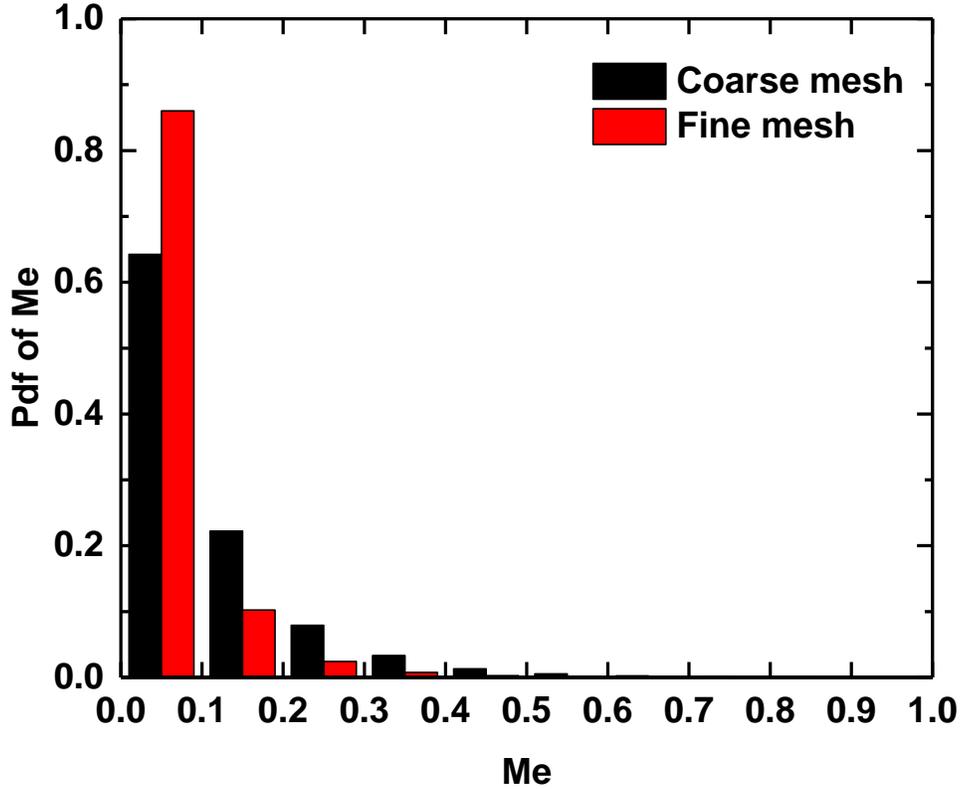

Figure. A5 Probability density function of $Me$ with different LES meshes.